\documentclass[preprint,11pt]{elsarticle}

\usepackage{graphicx}
\usepackage{amssymb}
\usepackage{dsfont}
\usepackage{amsmath}
\usepackage{mathdots}

\begin{document}

\begin{frontmatter}


\title{Causal structure and algebraic classification of\\ area metric spacetimes in four dimensions}
\author{Frederic P. Schuller}
\address{Max Planck Institute for Gravitational Physics, Albert Einstein Institute, Am M\"uhlenberg 1, 14476 Potsdam, Germany}
\ead{fps@aei.mpg.de}

\author{Christof Witte}
\address{Institut f\"ur Physik, Humboldt-Universit\"at zu Berlin, Newtonstrasse 15, 12489 Berlin, Germany}
\ead{witte@physik.hu-berlin.de}

\author{Mattias N. R. Wohlfarth}
\address{Zentrum f\"ur Mathematische Physik und II. Institut f\"ur Physik, Universit\"at Hamburg, Luruper Chaussee 149, 22761 Hamburg, Germany}
\ead{mattias.wohlfarth@desy.de}




\begin{abstract}
Area metric manifolds emerge as a refinement of symplectic and metric geometry in four dimensions, where in numerous situations of physical interest they feature as effective matter backgrounds. In this article,  this prompts us to identify those area metric manifolds that qualify as viable spacetime backgrounds in the first place, in so far as they support causally propagating matter.
This includes an identification of the timelike future cones and their duals associated to an area metric geometry, and thus paves the ground for a discussion of the related local and global causal structure in standard fashion.
In order to provide simple algebraic criteria for an area metric manifold to present a consistent spacetime structure, we develop a complete algebraic classification of area metric tensors up to general transformations of frame. Remarkably, a suitable coarsening of this classification allows to prove a theorem excluding the majority of algebraic classes of area metrics as viable spacetimes.
\end{abstract}






\end{frontmatter}


\newcommand{\winkel}{<\hspace{-1.25ex})\hspace{0.5ex}}
\newtheorem{definition}{Definition}[section]
\newtheorem{theorem}{Theorem}[section]
\newtheorem{lemma}{Lemma}[section]

\section{Introduction}
The assumption that physical spacetime features the structure of a Lo\-rentzian manifold \cite{Beem} impacts modern physical theory in at least two major ways. First, it conveniently restricts the admissible types of matter fields and their dynamics, as was first pointed out by Wigner \cite{Wigner} and is routinely used in particle physics \cite{EPP}. Indeed, the very idea of a Lorentzian spacetime geometry historically roots in the structure of Maxwell electrodynamics \cite{Einstein}. Second, dynamics for Lorentzian metrics with a well-posed initial value problem almost inevitably \cite{Kuchar} are those of Einstein-Hilbert theory, with the well-known physical implications \cite{MTW}: the big bang singularity, precession of planetary orbits, gravitational lensing and an expanding universe. Together, remarkably much of what we infer about the structure of spacetime, its matter contents, and indeed their interplay, hinges on the presumed Lorentzian spacetime structure.

However, particularly over the last decade, disturbingly robust and diverse observational evidence has been accumulated that there is something significant we currently do not understand about the matter contents of the universe, gravitational dynamics, or both. For instance, in order to explain the observed late-time accelerated expansion of the universe \cite{Spergel} and at the same time the data collected from the lensing of light through galaxies \cite{Knop}, one would have to assume that a spectacular 74\% of energy and 22\% of matter in the universe are of entirely unknown origin, and do not interact in any other conceivable way than gravitationally \cite{DEMrev}.

But while the existence of such vast amounts of dark energy and dark matter may indeed be the correct conclusion to be drawn from the observational data, this seems not an uncontestably plausible or compelling conclusion. Much less so because there are a number of further anomalies in gravitational physics, such as the flattened galaxy rotation curves, the anomalous accelerations of Pioneer 10 and 11, the fly-by anomaly, and others \cite{Laemmerzahl}. In summary, there is an increasing list of discrepancies between observation and theory, which in some cases hint at new particle physics \cite{newEPP}, in other cases at new gravitational physics \cite{MOND}, and one may well speculate that some hint at both \cite{Punzi:2006hy}.

Now on the one hand, it may be the case that all of these anomalies are mutually independent, and require a different resolution each. On the other hand, and this is the line of thought we want to pursue here, one would expect both, new gravitational and new particle physics, if the geometry of spacetime turned out to be different from that of a Lorentzian metric manifold. Such a generalized spacetime geometry would have to be sufficiently general to capture various of the anomalies currently escaping explanation, while at the same time providing feasible spacetime backgrounds for particle physics.

Area metric manifolds \cite{Schuller:2005yt} present such a promising candidate for a refinement of Lorentzian geometry. Indeed they arise as effective backgrounds in quantum electrodynamics \cite{Drummond:1979pp} and string theory \cite{Schuller:2005ru}, and present a considerable refinement of metric geometry in four dimensions. They already proved their worth in gravitational physics, both within standard general relativity \cite{Punzi:2009ks} and beyond \cite{Punzi:2006nx}, \cite{BDgeometry}. A comprehensive treatment of their properties as physically viable spacetimes in which matter can causally propagate, however, was not seriously attempted before.

In the present article we address this issue and study local and global conditions for four-dimensional area metric manifolds to provide a spacetime geometry. To this end, we first investigate the causal evolution of matter fields in order to define a causal structure on an area metric manifold, and study causality conditions of different strengths for area metric manifolds. Second, we give a complete overview of four-dimensional area metrics, which technically amounts to an algebraic classification of the area metric tensor. We obtain the remarkable result that the causality theory of area metric manifolds renders a great many of the possible algebraic classes we obtain as unphysical. Thus, for further considerations of area metric manifolds one may focus on only those algebraic classes that present physically viable spacetimes.

\vspace{0.5cm}
\textit{Outline.} In section \ref{chap:2}, we start by reviewing the most important aspects of area metric geometry, in so far as they play a role in the present work.  To get a feeling for the way in which area metric geometry presents a refinement of metric geometry, we investigate the mathematical properties of low-dimensional area metrics in section \ref{sec:22}, and their emergence from fundamental physics, such as quantum electrodynamics, in section \ref{sec:23}. Provided with the mathematical definitions, and supported by the obtained physical intuition for area metric geometry, we then turn to one of the key points of this article, namely we employ Maxwell theory to define a causal structure on area metric manifolds in section \ref{sec:24}. With the result of this construction, we are then equipped to present the central definitions of weakly and strongly hyperbolic area metric spacetimes in section \ref{sec:25}. These provide an \textit{analytic} characterization of area metric manifolds that present viable spacetime structures. Relevant global causality conditions for area metric manifolds can then be imposed in addition, and we observe that celebrated theorems, such as the equivalence of the Alexandrov topology with the underlying manifold topology, directly extend to area metric spacetimes.

In section \ref{chap:3}, we set aside the analytic considerations for a moment, and turn to an \textit{algebraic} classification of four-dimensional area metric manifolds. We obtain a complete overview of all possible area metric manifolds and provide, as a corollary to our classification theorem, a list of area metric normal forms. The provision of such normal forms, together with our detailed study of the respective algebras describing the involved gauge ambiguity, constitutes an immensely useful calculational tool, very much like in the familiar case of pseudo-Riemannian metrics.

We combine our findings on the analytic characterization of the causal properties of area metric spacetimes with the algebraic classification of four-dimensional area metrics in section \ref{chap:4}. This culminates in the proof that a large number of algebraic classes do not present area metric spacetimes. An even stronger version of this theorem can be obtained by focusing on phenomenologically important cases of highly symmetric area metric spacetimes.

In a conclusion, we finally place our results and the methods employed in this article in a wider context, emphasize what has been achieved and where the limitations of the current study lie.

\section{Area metric geometry and causal structure}
\label{chap:2}
The central aspects of area metric geometry, as far as they play a role for the developments in this article, are presented and discussed in this section. As an immediate physical question the causal structure of Maxwell theory on generic area metric manifolds is studied in some detail. These constructions culminate in the definition of strongly hyperbolic area metric spacetimes and present the first technical pillar of this article.

\subsection{Area metric manifolds}
\label{sub:21}
We start with the fundamental definitions of area metric geometry \cite{Punzi:2006nx} in $d$ dimensions which presents a generalization of metric geometry.
  \begin{definition}
     An area metric manifold (M,G) is a smooth d-dimensional manifold $M$ equipped with a fourth-rank covariant tensor field G with the following symmetry and invertibility properties at each point p of M:
     \renewcommand{\labelenumi}{(\roman{enumi})}
     \begin{enumerate}
     \item $G(X,Y,A,B)=G(A,B,X,Y)$ for all X,Y,A,B in $T_pM$
     \item $G(X,Y,A,B)=-G(Y,X,A,B)$ for all X,Y,A,B in $T_pM$
     \item For each $p$ of $M$ and $X,Y,A,B$ in $T_pM$ the map  $\hat{G}:\Lambda^2T_pM\rightarrow\Lambda^2T^*_pM$, defined through $\hat{G}(X\wedge Y)(A\wedge B):=G(X,Y,A,B)$ by linear continuation, is invertible. Its inverse then defines a fourth-rank contravariant tensor field $G^{-1}$ called the inverse area metric.
     \end{enumerate}
  \end{definition}
Here $\Lambda^2T_pM=T_pM\wedge T_pM$ denotes the space of all contravariant antisymmetric tensors of rank two and we will drop the hat on $G$ where no confusion arises.

Given a basis $\{e_a\}$ on $T_pM$, the symmetry conditions can be written in terms of the components $G(e_a,e_b,e_c,e_d)=G_{abcd}$ of the area metric:
\begin{equation}
G_{abcd}=G_{cdab}=-G_{bacd}.
\end{equation}
Due to these symmetries, the indices of $G$ may be combined to antisymmetric Petrov pairs $\lbrack ab \rbrack$ such that $G$ can be represented by a symmetric square matrix of dimension $D=d(d-1)/2$. More precisely, we introduce Petrov indices $A=1,\ldots,d(d-1)/2$ for every antisymmetric pair of small indices $[ab]$. The Petrov indices can be calculated as follows: without loss of generality we assume $a<b$ and calculate the Petrov index $A$ in terms of $a$ and $b$ as $A=(a(2d-3)-a^2)/2+b$. If it is not clear from the indices that we use the Petrov notation of an object $\Gamma$ we write $Petrov(\Gamma)$. In four dimensions for instance, which is the case of direct physical interest, we have index pairs $[01]$, $[02]$, $[03]$, $[12]$, $[31]$, $[23]$ with the corresponding Petrov indices $A=1,\ldots,6$. The independent components of an area metric $G$ in four dimensions may hence be arranged as the $6\times 6$ Petrov matrix
\begin{equation}
Petrov(G)=\left[\begin{array}{cccccc} G_{0101}&G_{0102} & G_{0103}&G_{0112}&G_{0131}&G_{0123}\\
                                       &G_{0202} & G_{0203}&G_{0212}&G_{0231}&G_{0223}\\
 		                             \ddots&         & G_{0303}&G_{0312}&G_{0331}&G_{0323}\\
 		                                   &\ddots   &         &G_{1212}&G_{1231}&G_{1223}\\
 		                                   &         & \ddots  &        &G_{3131}&G_{3123}\\
 		                                   &         &         & \ddots &       &G_{2323}
\end{array}\right]
\label{eqn:matrix}
\end{equation}
with the components under the diagonal filled by symmetry.

Some care has to be taken when extending the summation convention to Petrov indices. Since the summation over Petrov indices $A$ essentially corresponds to a sum over \textit{ordered} antisymmetric pairs of tangent space indices, we need to multiply by a factor of $1/2$ when resolving a contraction over Petrov indices in terms of a double contraction over an \textit{unordered} pair of tangent space indices: $X^A\Omega_A=1/2X^{ab}\Omega_{ab}$ for antisymmetric tensor fields $X$ and $\Omega$ of valence $(2,0)$ and $(0,2)$, respectively.

The invertibility requirement (iii) implies that the Petrov matrix $Petrov(G)$ representing $G$ is non-degenerate and $Petrov(G^{-1})=Petrov(G)^{-1}$. By direct calculation, we obtain that the components of the inverse area metric $G^{-1}$ satisfy the identity
\begin{equation}
(G^{-1})^{abmn}G_{mncd}=4\delta^{\lbrack a}_c\delta^{b\rbrack}_d.
\label{eqn:inv}
\end{equation}
where the factor of 4 arises due to the use of the above described summation convention and weighted antisymmetrization

Area metric geometry is a refinement of metric geometry, insofar as every pseudo-Riemannian manifold is an area metric manifold, but not all area metric manifolds are induced from a metric one. Nevertheless, it is sometimes interesting to discuss the following special type of area metrics:

\begin{definition}
An area metric $G$ is said to be metric-induced if there exists a metric $g$ such that
\begin{equation}
G(X,Y,A,B)=g(X,A)g(Y,B)-g(X,B)g(Y,A).
\label{eqn:indG}
\end{equation}
\end{definition}
For an area metric $G_g$ induced in this fashion we have that $G_g(X,Y,X,Y)=g(X,X)g(Y,Y)\sin^2\lbrack\winkel(X,Y)\rbrack$ is the squared area of the parallelogram spanned by vectors $X,Y$ as measured in the underlying metric geometry. For later use we write the metric-induced area metric in components:
\begin{equation}
G_{abcd}=g_{ac}g_{bd}-g_{ad}g_{bc}
\label{eqn:Metind2}
\end{equation}
and by virtue of equation (\ref{eqn:inv}) we have
\begin{equation}
(G^{-1})^{abcd}=g^{ac}g^{bd}-g^{ad}g^{bc}.
\label{eqn:metindinv}
\end{equation}

Note that for metric-induced and generic area metrics alike, any pair of $SL(2,\mathds{R})$-related parallelograms $(X,Y)$ and $(\tilde{X},\tilde{Y})$, i.e. $\tilde{X}=aX+bY$ and $\tilde{Y}=cX+dY$ with $ad-bc=1$, have identical areas as measured by the area metric, $G(X,Y,X,Y)=G(\tilde{X},\tilde{Y},\tilde{X},\tilde{Y})$. Thus an area metric does not distinguish parallelograms $(X,Y)$ that describe the same oriented area $X\wedge Y$. This property, together with the reproduction of the familiar notion of area in the metric-induced case, justifies to call $G$ an area metric.\\
It should be noted that a generic area metric contains more algebraic degrees of freedom than a metric, starting from dimension four. This can be seen by counting the independent components of the symmetric $D\times D$ Petrov matrix representing the area metric, which amounts to $D(D+1)/2$ independent real numbers. The invertibility requirement does not further reduce this number since it is an open condition. Thus area metrics in dimensions 2, 3, 4 and 5 have 1, 6, 21 and 55 independent components, respectively.

An area metric $G$ naturally gives rise to a scalar density $|\mbox{det}(Petrov(G))|^{1/(2d-2)}$ of weight $+1$. That $\mbox{det}(Petrov(G))$ transforms as a density of weight $2d-2$ under a change of frame on the underlying $d$-dimensional manifold,
\begin{equation}
\mbox{det}(Petrov(G_{mnpq}T^{\lbrack m}_{~~~a}T^{n\rbrack}_{~~~b}T^{\lbrack p}_{~~~c}T^{q\rbrack}_{~~~d}))=\mbox{det}(T_{~~b}^a)^{2d-2}\mbox{det}(Petrov(G)),
\end{equation}
for a transformation matrix $T^{a}_{~~b}$, follows from the identity
\begin{equation}
\mbox{det}(Petrov(T^{\lbrack a}_{~~~c}T^{b\rbrack}_{~~~d}))=(\mbox{det}(T^{a}_{~~b}))^{d-1}\, ,
\label{eqn:TwedgeT}
\end{equation}
which deserves a\\
\textbf{Proof \cite{Tuschik}.} Consider a $d$-dimensional vector space $V$ and an automorphism $T:~V\rightarrow V$. We define the induced endomorphism $T\wedge T:~V\wedge V\rightarrow V\wedge V$ on the induced $d(d-1)/2$-dimensional vector space $V\wedge V$ as $(T\wedge T)(v\wedge w)=T(v)\wedge T(w)$ for vectors $v,w\in V$. Choose an arbitrary vector $e_1\in V$ and first assume that
\begin{equation}
e_1~~~\mbox{and}~~~e_{i+1}:=T(e_i),
\label{eqn:basis}
\end{equation}
for $i=1,\dots,d-1$ defines a basis for $V$. The case in which this assumption does not immediately hold is discussed further below. Clearly $T(e_d)=\sum_{i=1}^d c_i e_i$ for coefficients $c_i$, so that in the basis $\{e_a\}$, the $d\times d$ matrix representing $T$ takes the form
\begin{equation}
T=\left[
\begin{array}{ccccc}
0&0 &0 &0&c_1\\
1&0&0&0&c_2\\
0&\ddots &0& 0 & \vdots\\
0&0&\ddots &0 &\vdots\\
0&0& 0&1&c_d
\end{array}
\right]\, ,
\end{equation}
such that one recognizes that $\mbox{det}(T)=(-1)^{d-1}c_1$ and so we have $(\mbox{det}(T))^{d-1}=(-1)^{d-1}c_1^{d-1}$. Now do also construct the induced basis $\{e_a\wedge e_b\}$, with $a<b$, on $V\wedge V$, and choose the order
\begin{equation}
e_1\wedge e_2,\dots,~e_1\wedge e_d,~ e_2\wedge e_3,\dots,~e_2\wedge e_d,\dots,~e_{d-1}\wedge e_d\,.
\end{equation}
Using the definition of $T\wedge T$, we may now calculate the $d(d-1)/2$-dimensional square matrix representing $T\wedge T$ in this basis. In four dimensions for instance, the $6\times 6$ matrix representing $T\wedge T$ takes the form
\begin{equation}
T\wedge T=\left[
\begin{array}{cccccc}
0&0&-c_1&0&0&0\\
0&0&0&0&-c_1&0\\
0&0&0&0&0&-c_1\\
1&0&c_2&0&-c_2&0\\
0&1&c_4&0&0&-c_2\\
0&0&0&1&c_4&-c_3
\end{array}
\right]\,.
\end{equation}
We may then calculate the determinant $\mbox{det}(T\wedge T)$ by recursively expanding all required minors with respect to their first rows, say. Together with the choice of basis we made, this implies that after $d-1$ steps the remaining minor to calculate is the determinant of the $(d-1)(d-2)/2$-dimensional unit matrix. The result of this calculation is $\mbox{det}(T\wedge T)=(-1)^{(d-1)(d^2-2d+6)/3}(-c_1)^{d-1}$. Since by construction the exponent $(d-1)(d^2-2d+6)/3$ is an integer and its divisibility by $2$ is not affected by multiplication by 3, it is always an even integer. Thus we arrive at $\mbox{det}(T\wedge T)=(-1)^{d-1}(c_1)^{d-1}$, which under the assumption that (\ref{eqn:basis}) already defines a basis for $V$ concludes the proof. It remains to show that if the first $k<d$ basis vectors form an invariant subspace of $V$, i.e. $T(e_k)=\sum_{i=1}^k c_i e_i$ for some $k<d$, the identity (\ref{eqn:TwedgeT}) still holds. In this case we have to choose another arbitrary vector $e_{k+1}\in V$ that is linearly independent from the $e_i$ with $i\le k$ to construct the next basis vectors according to (\ref{eqn:basis}). Repeat this procedure until a complete basis is found. Then the matrices representing $T$ and $T\wedge T$ decompose into block-diagonal form and the determinant is separately taken over every block in the same fashion as shown above. This yields the same result, and completes the proof.\\

Employing the density $|\mbox{det}(Petrov(G))|^{1/(2d-2)}$, we can define a volume form $\omega_G$ on an area metric manifold:

\begin{definition}
An area metric manifold $(M,G)$ carries a canonical volume form $\omega_G$, defined by
\begin{equation}
{\omega_G}_{a_1\cdots a_d}=|\mbox{det}(Petrov(G))|^{1/(2d-2)}\epsilon_{a_1\cdots a_d},
\label{eqn:volf}
\end{equation}
where $\epsilon$ is the Levi-Civita tensor density normalized such that $\epsilon_{0\cdots d-1}=1$.
\end{definition}
The volume form plays an essential role in our algebraic classification of four-dimensional area metrics, as we will see in section 3.

Having introduced the very basic notions of area metric geometry we analyse low-dimensional area metric manifolds in some detail, in the next section. Apart from conveying some further intuition for area metrics, we will discuss some particular properties of area metrics in four dimensions, which presents the case of most immediate physical interest for this article.

\subsection{Low dimensional area metric manifolds}
\label{sec:22}
The study of low-dimensional cases of area metric manifolds reveals two insights. On the one hand it illustrates in what sense area metrics are a refinement of metric geometry. On the other hand we will see that in four dimensions area metrics play a very special role indeed.
\renewcommand{\labelenumi}{$d=$\arabic{enumi}:}

\begin{enumerate}
\item There are no area metrics in one dimension. For from the symmetries of the area metric tensor $G$ it is clear that there is no non-vanishing component of the area metric in only one dimension. Thus no such $G$ can be invertible.

\item In two dimensions an area metric $G$ is entirely determined by a scalar density $\tilde\Phi=G_{abcd}\epsilon^{ab}\epsilon^{cd}/4$ of weight $+2$ by virtue of
\begin{equation}
G_{abcd}=\tilde\Phi(\epsilon_{ac}\epsilon_{bd}-\epsilon_{ad}\epsilon_{bc}),
\end{equation}
where $\epsilon_{ab}$ denotes the components of the totally antisymmetric tensor density. This can be seen by contracting both sides with $\epsilon^{ab}\epsilon^{cd}$. The only remaining component of the area metric tensor is $G_{0101}=\tilde\Phi$ and all other unrelated components vanish. Thus in two dimensions, an area metric is not a refinement of a metric, but rather a coarser structure. In fact, area metric geometry in two dimensions is symplectic geometry \cite{Symplectic} with the symplectic form $\tilde\Phi^{1/2}\epsilon$.

\item An area metric in three dimensions has six independent components, just like a metric. This is more than a coincidence. We can even show that every area metric $G$ in three dimensions is metric-induced, with the inducing metric
\begin{equation}
g_{ab}=\frac{1}{8}\omega_G^{ijk}\omega_G^{pqr}G_{arij}G_{pqkb}.
\label{eqn:gfromG}
\end{equation}
Indeed, one easily verifies that $g_{ab}=g_{ba}$ and
\begin{equation}
0\not=\mbox{det}(Petrov(G))=\mbox{det}(Petrov(g_{a\lbrack c}g_{d\rbrack b}))=(\mbox{det}~g)^{d-1},
\end{equation}
again using the identity (\ref{eqn:TwedgeT}), proves that $g$ is indeed a metric. What remains to be shown is that the area metric $G$ is in fact induced by this metric $g$. For that purpose we write the area metric in Petrov notation
\begin{equation}
Petrov(G)=\left[\begin{array}{ccc} G_{0101}&G_{0102} & G_{0112}\\
                           G_{0102} & G_{0202}&G_{0212}\\
 		                       G_{0112}&G_{0212}&G_{1212}
 		
\end{array}\right]
\label{eqn:G3D}
\end{equation}
and the components of the inverse area metric volume form $\omega_G$ read
\begin{equation}
\omega_G^{ijk}=|\mbox{det}(Petrov(G))|^{-1/4}\epsilon^{ijk},
\end{equation}
where the determinant is taken over the matrix (\ref{eqn:G3D}). We now show the proposition for the component $G_{0101}$ of the area metric. We need to calculate the components $g_{00}$, $g_{11}$ and $g_{01}$. According to equation (\ref{eqn:gfromG}) these are
\begin{eqnarray*}
g_{00}&=&|\mbox{det}(Petrov(G))|^{-1/2}(G_{0101}G_{0202}-G_{0102}G_{0102}),\\
g_{11}&=&|\mbox{det}(Petrov(G))|^{-1/2}(G_{0101}G_{1212}-G_{0112}G_{0112}),\\
g_{01}&=&|\mbox{det}(Petrov(G))|^{-1/2}(G_{0212}G_{0101}-G_{0102}G_{0112}).
\end{eqnarray*}
Inserting this into equation (\ref{eqn:Metind2}) and using the determinant of the matrix (\ref{eqn:G3D}) proofs the equality
\begin{equation}
g_{00}g_{11}-(g_{01})^2=G_{0101}.
\end{equation}
Repeating this calculation for the other components of $G$ completes the proof. This means area metric geometry in three dimensions is metric geometry, and vice versa. Thus the three-dimensional area metric geometry may be viewed as metric or area metric, with no way to distinguish one from the other. This result is implicit in Cartan's treatise \cite{Cartan}

\item In four dimensions, an area metric has 21 independent components, whereas a metric has only 10. Thus an area metric contains more algebraic degrees of freedom than a metric. It is intuitively clear that using a $GL(4)$ transformation, at most $16$ of the $21$ parameters of the area metric at a point can be brought to zero. A generic area metric can therefore be expected to locally determine up to five $GL(4)$-scalars. That this is indeed the case will be an essential result in the algebraic classification in section 3. This classification will rely on the remarkable feature that in four dimensions, the canonical volume form defined by (\ref{eqn:volf}) is an area metric in its own right.
\end{enumerate}

One may justifiedly wonder whether one could consider even more refined structures than an area metric, such as a 3-volume metric $V_{\lbrack abc\rbrack\lbrack def\rbrack}$, a 4-volume metric, and so on. However, while a 3-volume metric would indeed be a refinement of area metric geometry on manifolds of dimension six or higher, one easily verifies that in dimension four, a 3-volume geometry is actually coarser than an area metric geometry. Essentially this is clear by dualizing the antisymmetric triple $\lbrack abc\rbrack$ using the volume form. Similarly for higher forms in higher dimensions. In this sense area metric geometry is the most refined geometric structure in the above sequence, when we consider the physically immediately relevant case of four dimensions.

The four examples presented here shall be sufficient to get a feeling for how area metric geometry differs from metric geometry. Area metric manifolds emerge in various contexts in fundamental physics. This is illustrated by three examples in the next section.

\subsection{Emergence of area metric manifolds in fundamental physics}
\label{sec:23}
Surprisingly, area metrics naturally emerge in standard physical theory. Roughly speaking, area metrics appear as effective backgrounds upon the quantisation of matter. Let us make this more precise for the example of Maxwell electrodynamics and string theory. We also hint at applications to more speculative theories such as those featuring a non-symmetric spacetime metric. In the following we set $c=\hbar=1$.

\vspace{0.5cm}
\textit{Maxwell electrodynamics.} Consider the action for a classical electromagnetic field on a curved spacetime,
\begin{equation}
S[A]=-\frac{1}{4}\int d^4x\sqrt{-g}F_{ab}F_{cd}g^{ac}g^{bd},
\label{eqn:Elec}
\end{equation}
in terms of the one-form potential $A$ and the spacetime metric $g$, where $F=dA$ denotes the electromagnetic field strength. Explicitly using the antisymmetry of $F$, one may rewrite the action, fully equivalently, in area metric form
\begin{equation}
S[A]=-\frac{1}{8}\int d^4x|\mbox{det}(Petrov(G))|^{1/6}F_{ab}F_{cd}(G^{-1})^{abcd},
\label{eqn:ElecMetind}
\end{equation}
where $G$ is the area metric induced by a metric $g$ according to (\ref{eqn:metindinv}). In the following we write $G^{abcd}$ for $(G^{-1})^{abcd}$.\\
However it is obvious that one could easily consider Maxwell theory on a generic (rather than metric-induced) area metric manifold. One could, but why would one? The answer is provided by consideration of the quantum theory corresponding to (\ref{eqn:Elec}). Drummond and Hathrell calculated the one-loop effective action for photon propagation, on a curved background, by taking into account the production of virtual electron-positron pairs in the framework of quantum electrodynamics \cite{Drummond:1979pp}. In a gravitational vacuum (i.e., where Lorentzian spacetime is Ricci flat) they obtain the effective electromagnetic field action
\begin{equation}
W\lbrack A\rbrack\sim\int d^4x\sqrt{-g}(g^{a\lbrack c}g^{d\rbrack b}+\lambda C^{abcd})F_{ab}F_{cd}+O(\lambda^2),
\label{eqn:DHA}
\end{equation}
where $C$ denotes the Weyl curvature tensor of $g$ and $\lambda=\alpha/(90\pi m^2)\cong 3,85~ \mbox{fm}^2$ (with electron mass $m$ and fine structure constant $\alpha$) is the characteristic scale of the interaction. To leading order in $\lambda$, one may view this as classical Maxwell theory of the form (\ref{eqn:ElecMetind}) on an area metric manifold with area metric $G_{DH}=g^{a\lbrack c}g^{d\rbrack b}+\lambda C^{abcd}$. Thus the first order quantum corrections of Maxwell theory in vacuum on a curved spacetime may be absorbed into an area metric structure \cite{Punzi:2009ks}. One cannot decide whether one is dealing with first order quantum effects of electrodynamics on a metric spacetime, or classical electrodynamics on an area metric manifold.

\vspace{0.5cm}
\textit{String theory.} As a second example, we consider the Nambu-Goto action for an open string \cite{Polch:1} moving through a target metric manifold $(M,g)$ of dimension $d$. We parametrize the worldsheet by two parameters $\sigma$ and $\tau$ so that the worldsheet area is determined by the pull-back of the metric $g$ to the worldsheet through
\begin{equation}
S_{NG}\lbrack x\rbrack=\int d\sigma d\tau \sqrt{\mbox{det}(\partial_{\alpha}x^a\partial_{\beta}x^bg_{ab}(x))},
\label{eqn:Nambu}
\end{equation}
where the partial derivatives denote differentiation with respect to $\sigma$ and $\tau$ and $x^a$ denotes the embedding functions of the worldsheet. Using the metric-induced area metric $G$ defined in (\ref{eqn:indG}) the Nambu-Goto action takes the form
\begin{equation}
S_{NG}\lbrack x\rbrack=\int d^2\sigma\sqrt{(G)_{abcd}(x)\dot{x}^ax'^b\dot{x}^cx'^d},
\end{equation}
where the prime and the dot denote differentiation with respect to $\sigma$ and $\tau$, respectively. Again, the action could be immediately generalized by replacing the metric induced area metric with a general area metric G. But as in the case of electrodynamics, there is no compelling reason for doing so in the classical theory. However, quantisation again teaches us otherwise. The Nambu-Goto action (\ref{eqn:Nambu}) itself is difficult to quantize because of the non-linearity produced by the square root. One circumvents this problem by consideration of the classically equivalent Polyakov action \cite{Polyakov}
\begin{equation}
S_P\lbrack x, \gamma\rbrack=\frac{1}{2}\int d\sigma d\tau \sqrt{-\gamma}\gamma^{\alpha\beta}\partial_{\alpha}x^a\partial_{\beta}x^bg_{ab}\,.
\label{eqn:Poly1}
\end{equation}
Here $\gamma$ is an independent two-dimensional worldsheet metric. That the Polyakov action is classically equivalent to the Nambu-Goto action, one sees by using the Euler-Lagrange-equation for the worldsheet metric $\gamma$ to eliminate the latter from the action (\ref{eqn:Poly1}). Now in the presence of a highly excited (coherent) quantum state \cite{Polch:1}, an individual string effectively perceives a generalized background defined by a metric $g_{ab}$ and an antisymmetric two-form $B_{ab}$ (for the purpose of this illustration we disregard the dilaton field),
\begin{equation}
S_P\lbrack x, \gamma\rbrack=\frac{1}{2}\int d\sigma d\tau \sqrt{-\mbox{det}\gamma}(\gamma+\epsilon)^{-1~\alpha\beta}\partial_{\alpha}x^a\partial_{\beta}x^b(g_{ab}+B_{ab})\,.
\label{eqn:Poly2}
\end{equation}
Again, $\gamma$ is an independent worldsheet metric and $\epsilon_{\alpha\beta}$ the totally antisymmetric tensor density on the worldsheet. To show how this action (\ref{eqn:Poly2}) gives rise to an area metric background, we rewrite it in the form \cite{Schuller:2005ru}
\begin{equation}
S_P\lbrack x, \gamma,\lambda\rbrack=\frac{1}{2}\int d\sigma d\tau \sqrt{-(1-\lambda^2)\mbox{det}(\gamma-\lambda\epsilon)}(\gamma-\lambda\epsilon)^{-1~\alpha\beta}\partial_{\alpha}x^a\partial_{\beta}x^b(g_{ab}+\lambda^{-1}B_{ab})\, ,
\label{eqn:Poly3}
\end{equation}
introducing an auxiliary scalar $\lambda$. Expanding $\mbox{det}(\gamma-\lambda\epsilon)$ and $(\gamma-\lambda\epsilon)^{-1}$ with respect to $\lambda$ shows the equivalence with (\ref{eqn:Poly2}). This formally looks like a Polyakov action with a non-symmetric spacetime metric $g+\lambda^{-1}B$ which, however, has no clear geometric interpretation. Similarly, $\tilde\gamma=\gamma-\lambda\epsilon$ can be considered as a non-symmetric worldsheet metric, which can be eliminated as usual, leading to the Nambu-Goto type action
\begin{equation}
S_{NG}\lbrack x,\lambda\rbrack=\int\sqrt{-(1-\lambda^2)\mbox{det}(\partial_{\alpha}x^a\partial_{\beta}x^b(g_{ab}+\lambda^{-1}B_{ab})(x))}.
\label{eqn:ngmb}
\end{equation}
Now this Nambu-Goto action has a clear interpretation in area metric geometry: for any invertible map $m:~T_pM\rightarrow T_p^*M$, we introduce an area metric $G_m$ by
\begin{equation}
(G_m)_{abcd}=\frac{1}{2}(m_{ac}m_{bd}-m_{ad}m_{bc}+m_{ca}m_{db}-m_{cb}m_{da}).
\label{eqn:symAM}
\end{equation}
Then equation (\ref{eqn:ngmb}) may be written as
\begin{equation}
S_{NG}\lbrack x,\lambda\rbrack=\int\sqrt{(1-\lambda^2)G_{g+\lambda^{-1}B}(\dot x,x',\dot x,x')}.
\label{eqn:ngG}
\end{equation}
This shows that the motion of a string on a general background may be effectively described as the motion on an area metric spacetime by means of the action (\ref{eqn:ngG}).

\vspace{0.5cm}
\textit{Non-symmetric gravity theory.} A spacetime geometry described by a non-sym\-metric metric was already proposed by Einstein \cite{Einstein:1}, and the idea was followed up in various forms by many others (see for example \cite{Papapetrou:1948jw}, \cite{Moffat:1995fc}). In all incarnations, the theory is based on a non-degenerate bilinear form $m_{ab}$ that has both a symmetric and non-symmetric part.\\
One of the fundamental physical questions when proposing such a theory is how to consistently choose matter couplings. Rather than postulating an ad-hoc point particle action, it is prudent to ensure inner-theoretical consistency by starting with some more fundamental theory, such as Maxwell electrodynamics. The motion of light rays may then be derived, rather then stipulated. As far as we are aware, this approach has not been followed before when deciding on point-particle couplings to non-symmetric metric backgrounds. More precisely, one may choose the extension of Maxwell theory to non-symmetric backgrounds to be governed by the action (\ref{eqn:ElecMetind}) with the area metric defined by (\ref{eqn:symAM}) in terms of the non-symmetric field $m$. A careful analysis shows that also the motion of massive test particles \cite{Punzi:2009yq} is then described by the same geometric structure that governs the propagation of light in the geometric optical limit of Maxwell theory \cite{Punzi:2007di}. We will see in the next section that this geometric structure is indeed entirely determined by the area metric. Thus the propagation of test particles through a manifold is determined, once the coupling of electrodynamics is specified. In this fashion, all findings of this article immediately apply also to the study of non-symmetric backgrounds.

With the above list of examples for the emergence of area metric tensors from fundamental physics in mind, we now start addressing the crucial issue of the causal structure defined by such backgrounds, which will finally lead to analytic criteria underpinning our definition of area metric spacetimes.

\subsection{Causal structure of area metric manifolds}\label{sec:24}
In the following we want to study the causal structure of four-dimensional area metric manifolds. To this end we have to carefully consider what causality means and how statements about the causal structure of area metric manifolds can be deduced from first principles.

It is very important to understand that causality is not an intrinsic property of an underlying background geometry, but rather the effect of an interplay between fundamental matter propagating on the manifold and the geometric properties of the latter \cite{Laemmerzahl2}. That is, the analysis of the causal structure of any manifold is deeply related to the causal analysis of the field equations that govern the evolution of some particular matter field on the manifold. For example, Maxwell electrodynamics on a metric manifold has a well-posed initial value problem only if the metric has Lorentzian signature \cite{Wald}.

Precisely the same way, one may employ Maxwell electrodynamics on a generic four-dimensional area metric manifold $(M,G)$, see \cite{Schuller:2005yt},
\begin{equation}
S[A]=-\frac{1}{8}\int d^4x|\mbox{det}(Petrov(G))|^{1/6}F_{ab}F_{cd}G^{abcd}
\end{equation}
with $F=dA$, in order to define a causal structure, and this is what we will do here. The definition of the field strength $F$ in terms of a gauge potential $A$ and variation of the above action with respect to the latter lead to the electromagnetic equations of motion for the field strength $F$ and induction $H$,
\begin{equation}
dF=0~~~\mbox{and}~~~dH=0,
\label{eqn:fieldeqn1}
\end{equation}
with the electromagnetic induction $H$ being related to the field strength $F$ through
\begin{equation}
H_{ab}=-\frac{1}{4}|\mbox{det}(Petrov(G))|^{1/6}\epsilon_{abmn}G^{mnpq}F_{pq}\,.
\label{eqn:constRel}
\end{equation}

The causal structure of Maxwell theory on an area metric manifold is of course fully contained in the field equations (\ref{eqn:fieldeqn1}), which may be written in components as
\begin{equation}
(\omega_G^{-1})^{abcd}\partial_b F_{cd}=0,~~~|\mbox{det}(Petrov(G))|^{-1/6}\partial_b(|\mbox{det}(Petrov(G))|^{1/6}G^{abcd}F_{cd})=0,
\label{eqn:fieldeqn}
\end{equation}
with the inverse area metric volume form $\omega_G^{-1}$ defined according to (\ref{eqn:inv}), which is applicable since in four dimensions the volume form (\ref{eqn:volf}) is itself an area metric.

For a complete description of the initial value problem of Maxwell electrodynamics we further have to specify initial data. We introduce coordinates $x^a=(t,x^{\alpha})$ such that our initial data surface $\Sigma$ is described by $t=0$ and we define the electric and magnetic fields as $E_{\alpha}=F(\partial_t,\partial_{\alpha})$ and $B^{\alpha}=\omega_G^{-1}(dt,dx^{\alpha},F)$, respectively.
Now observe that the system (\ref{eqn:fieldeqn}) provides eight equations for six fields $(E_{\alpha},B^{\alpha})$, however, two of these eight equations are constraint equations. Indeed, in the chosen coordinates, the $t$-components of the two equations (\ref{eqn:fieldeqn}) do not contain any time derivatives:
\begin{eqnarray}
C_1&=&(\omega_G^{-1})^{0bcd}\partial_b F_{cd}=0,\\
C_2&=&|\mbox{det}(Petrov(G))|^{-1/6}\partial_b(|\mbox{det}(Petrov(G))|^{1/6}G^{0bcd}F_{cd})=0\,.
\label{eqn:constraints}
\end{eqnarray}
Thus they constrain the initial data one may provide for the fields $(E_{\alpha},B^{\alpha})$. Using the remaining evolution equations one finds that
\begin{equation}
\partial_t C_{1,2}=-C_{1,2}\partial_t\ln|det(Petrov(G))|^{1/6},
\end{equation}
so that the constraints are preserved under evolution in time.

The evolution equations themselves now are of the general form
\begin{equation}
{A^b}^M_{~~N}\partial_b u^N+B^M_{~~N}u^N = 0,
\label{eqn:diffeq}
\end{equation}
where $u^N=(E_{\alpha},B^{\alpha})$ and the four $6\times 6$ matrices $A^b$
\begin{equation}
A^0=\left[\begin{array}{cc}
G^{0\mu 0\nu} &0\\
0 & \delta^{\mu}_{\nu}
\end{array}\right],~~~
A^{\alpha}=\left[\begin{array}{cc}
-2 G^{0(\mu \nu)\alpha} &-\frac{1}{2}(\omega_G)_{0\nu\gamma\delta}G^{\gamma\delta\mu\alpha}\\
(\omega_G^{-1})^{0\mu\nu\alpha} & 0
\end{array}\right].
\end{equation}

From the theory of partial differential equations \cite{PDE:1}, it is known that the local causal behaviour of such a system of differential equations is encoded in the so-called characteristic polynomial $P(k)=\mbox{det}(A^bk_b)$ defined over transversal ($k_0\not= 0$) covectors $k$. Casting this expression into manifestly covariant form (conveniently rescaling $k_0$ to be unity), one finds for a four-dimensional area metric manifold
\begin{equation}
P(k)=-|\mbox{det}(Petrov(G))|^{-1/3}\mathcal G(k,k,k,k),
\label{eqn:principal}
\end{equation}
where the quartic Fresnel polynomial $\mathcal G(k,k,k,k)$ is defined as
\begin{equation}
\mathcal G(k,k,k,k)=-\frac{1}{24}(\omega_G)_{mnpq}(\omega_G)_{rstu}G^{mnr(a}G^{b|ps|c}G^{d)qtu}k_ak_bk_ck_d\,.
\label{eqn:fresnelpoly}
\end{equation}

The tensor $\mathcal{G}$ and its physical interpretation has been first obtained by Rubilar \cite{Rubilar} in the context of pre-metric electrodynamics \cite{Hehlbook}, by studying the propagation of electromagnetic field discontinuities. Our derivation here is a complementary one, which we choose since it directly leads to the related causality theory.

Furthermore, the theory of partial differential equations shows that a necessary condition for electromagnetic fields to propagate through an area metric manifold at all, is that the characteristic polynomial (\ref{eqn:principal}) admits non-vanishing null covectors, $P(k)=0$. Using the definition of the characteristic polynomial this condition reduces to the Fresnel equation
\begin{equation}
\mathcal G(k,k,k,k)=0.
\label{eqn:fresnel}
\end{equation}
From the linearity of the Fresnel polynomial, it follows that the null covectors constitute a cone $L_p$ in each cotangent space $T_p^*M$, i.e. a subset $L_p\subset T_p^*M$ such that $\lambda L_p\subseteq L_p$ for any real positive $\lambda$. Physically speaking, this statement on the admissible wave covectors $k$ is one on the geometric-optical limit of Maxwell theory.

\begin{figure}[t!]
	\centering
		\includegraphics[width=1\textwidth]{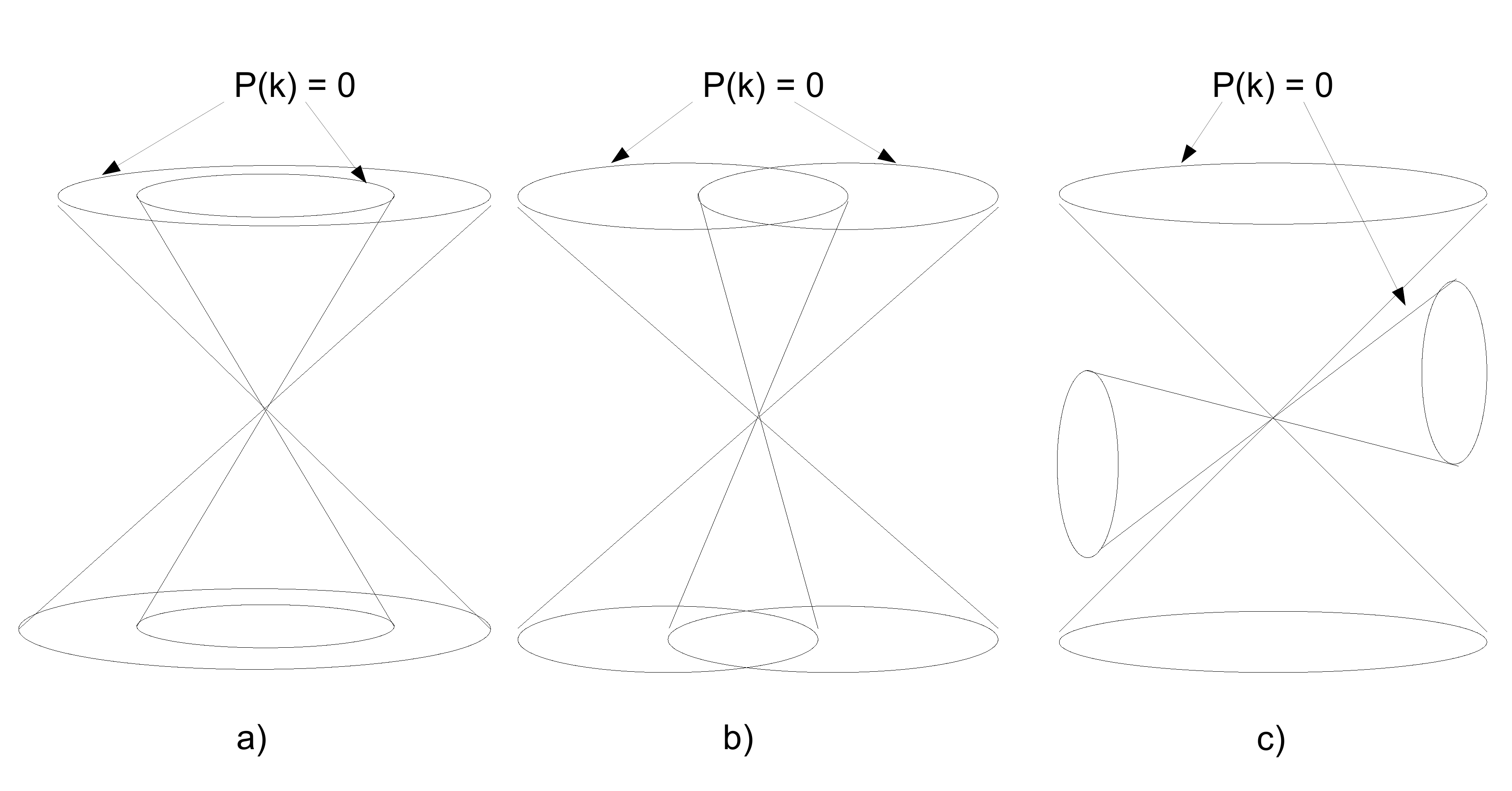}
		\caption{some quartic null cones in cotangent space}
	\label{fig:Nullstructure}
\end{figure}

For a metric-induced area metric (\ref{eqn:Metind2}), the quartic Fresnel equation (\ref{eqn:fresnel}) factorizes to the bi-quadratic equation $(g^{ab}k_ak_b)^2=0$, which in turn reproduces the familiar notion of covector null cones in Lorentzian geometry. However, the generic case of an area metric manifold leads to more elaborate local null structures (see figure \ref{fig:Nullstructure} for examples). Going beyond the geometric-optical limit of Maxwell theory one observes that the polarization of light determines which sheet of the surface in cotangent space defined by the quartic condition (\ref{eqn:fresnel}) is chosen \cite{Punzi:2007di}.

\subsection{Convex causal future cones and their duals}\label{sec:24a}
\begin{figure}[t!]
	\centering
		\includegraphics[width=1\textwidth]{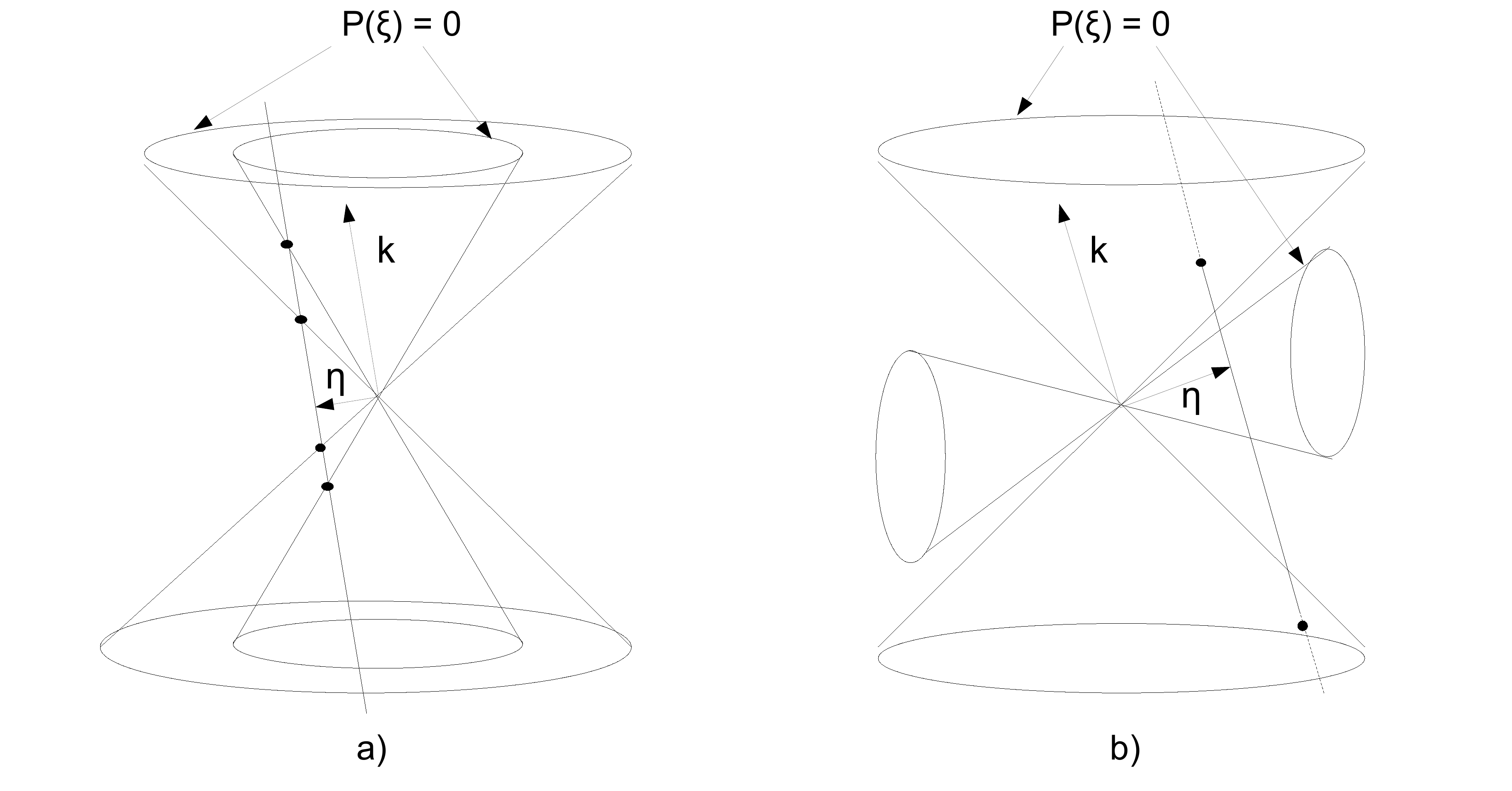}
		\caption{Illustration of a) a timelike covector $k$ and b) a covector $k$ that cannot be timelike}
	\label{fig:timelike}
\end{figure}

Once we ensured that the field propagates at all, we may turn to the question of well-posedness of the initial value problem for Maxwell theory. To this end we need to ensure that there are initial data surfaces $\Sigma$, which can only in the case if there are covectors $k$ normal to $\Sigma$, i.e. $k(\Sigma)=0$, and for which $P(k)\not=0$ and $P(\eta-\lambda k)$ has only real roots $\lambda$ for any covector $\eta$. Any such covector $k$ on a four-dimensional area metric manifold is called timelike. Geometrically, a covector $k$ is timelike if \textit{any} line in the direction of $k$ intersects the surface of null covectors four times. This is illustrated in figure \ref{fig:timelike}. We see that timelike covectors exist in the example from figure \ref{fig:Nullstructure}a but there are no timelike covectors in the example from figure \ref{fig:Nullstructure}c.

With the help of the characteristic polynomial $P(k)$ it is also possible to distinguish between future and past with respect to a given time orientation, which is chosen in terms of an everywhere timelike covector field $\tau$: we define the future timelike covector cone $C_p^*$ at a point $p\in M$ as the set of all covectors $\xi\in T_p^* M$ such that the roots $\lambda$ of $P(\eta-\lambda\xi)$ for any covector $\eta\in T_p^*M$ are positive with respect to the time orientation $\tau$. It can be shown that the future timelike covector cone is a convex cone \cite{Garding1}, \cite{Renegar}, i.e. for any covector $v\in C_p^*$ we have $\lambda v\in C_p^*$ for any $\lambda \in \mathds{R}^+$ and for any two covectors $v,w\in C_p^*$ it is true that $v+w\in C_p^*$. Furthermore, $C_p^*$ is open. These two properties and the fact that any two covectors $v,w\in C_p^*$ satisfy the inverse triangle inequality $P(v+w)\geq P(v)+P(w)$ \cite{Garding2} render the concept of the future timelike covector cone on an area metric manifold a true generalization of Lorentzian geometry.

It should be noted that once one has identified future timelike covectors it is also possible to define future timelike vectors, and to relate them in a one-to-one fashion. To this end, we define the cone $C_p$ of future timelike vectors at a point $p$, which itself is open and convex, as the set $C_p=\{v\in T_pM|k(v)\geq 0 ~\mbox{for all}~k\in C_p^*\}$, which is the dual cone to $C_p^*$. Note that a general feature of the relation between a cone and its dual is that given two cones $C_1^*$ and $C_2^*$ with $C_1^*\subset C_2^*$ we have $C_2\subset C_1$ for the dual cones $C_1$ and $C_2$. Thus there is an inversion of the inclusion relation when considering two cones and their respective dual cones.

It also turns out that the duality map between future timelike covectors and future timelike vectors is given in terms of the Fresnel tensor $\mathcal{G}$, whose components may be computed by differentiation of the Fresnel polynomial (\ref{eqn:fresnelpoly}) with respect to the components of the covector $k$.
\begin{theorem}\label{theo:21}
Let (M,G) be a four-dimensional area metric manifold. Let $C_p^*\subset T_p^*M$ be the future timelike covector cone at a point $p\in M$. Then for any $k\in C_p^*$ there is a bijection to the cone $C_p$ of future timelike vectors,
\begin{equation}
C_p^*\rightarrow C_p\, ,~~~k\mapsto \mathcal{G}(k,k,k,\cdot).
\end{equation}
\end{theorem}
\textbf{Proof.} The proof uses the fact that the mapping $C_p^*\rightarrow \mathds{R}$ defined through $\tau\mapsto -\ln\mathcal{G}(\tau,\tau,\tau,\tau)$ is the so-called self-concordant barrier functional. According to \cite{Renegar2}, see theorem 27 of \cite{Renegar}, it follows that the mapping $C_p^*\rightarrow C_p$ defined as $k\mapsto D\ln\mathcal{G}(k,k,k,k)$ is a bijection. Now it is easily checked that
\begin{equation}
(D\ln\mathcal{G}(k,k,k,k))(\tau)=\frac{\mbox{d}}{\mbox{dt}}\Big |_{t=0}\ln\mathcal{G}(k+t\tau,k+t\tau,k+t\tau,k+t\tau)=4\frac{\mathcal{G}(k,k,k,\tau)}{\mathcal{G}(k,k,k,k)}
\end{equation}
for any $\tau\in C_p^*$. Since the denominator never vanishes on the future timelike covector cone $C_p^*$, one finds that also $k\mapsto \mathcal{G}(k,k,k,\cdot)$ is a bijection $C_p^*\rightarrow C_p$, which concludes the proof.

\vspace{0.5cm}
Thus Theorem \ref{theo:21} describes the area metric spacetime analogue of raising and lowering indices on timelike vectors and covectors in Lorentzian geometry. It will also be useful to have the

\begin{definition}
Let $(M,G)$ be an area metric spacetime. A vector $X\in T_pM$ is called a future causal vector if $X$ lies in the closure $\overline{C_p}$ of the future timelike vector cone $C_p$.
\end{definition}

It should be noted that vectors $X$ lying on the boundary $\partial \overline{C_p}$ of the closure $\overline{C_p}$ are indeed null vectors, but (in contrast to the special case of Lorentzian metric manifolds) not every null vector lies on this boundary. We will have opportunity to return to this definitions when we discuss the global causal structure of area metric spacetimes. Note that (given a time orientation) past timelike vectors and past causal vectors may be defined accordingly.

\subsection{Area metric spacetimes}
\label{sec:25}

We now turn to the definitions of area metric spacetimes, which include conditions of various strength, beyond the mere area metric manifold structure.

\begin{definition}\label{def:25}
Let $(M,G)$ be a four-dimensional area metric manifold. We call $(M,G)$ a weakly hyperbolic area metric spacetime if there exists a time orientation in terms of an everywhere timelike covector field $\tau$.
\end{definition}
\renewcommand{\labelenumi}{\arabic{enumi}.}

The so defined weakly hyperbolic area metric spacetimes are only necessary, but not sufficient to ensure a well-posed initial value problem for Maxwell theory described by the action (\ref{eqn:ElecMetind}). Indeed, it will be instructive to formulate a notion of strongly hyperbolic area metric spacetimes, which in fact guarantees that the initial value problem for Maxwell theory is well posed, at least locally. To this end, we rewrite the first order PDE system (\ref{eqn:diffeq}) in the form
\begin{equation}
\partial_0 u^M={C^{\alpha}}^M_{~~N}\partial_{\alpha}u^N+D^M_{~~N}u^N,
\label{eqn:Cauchy}
\end{equation}
where $C^{\alpha}=-(A^{0})^{-1}A^{\alpha}$ and $D=-(A^{0})^{-1}B$. Now, consider the matrix $C(k)=C^{\alpha}k_{\alpha}$, where $k_{\alpha}$ are the components of some purely spatial covector $k$. The matrix $C(k)$ plays a key role in the definition of strongly hyperbolic area metric spacetimes.

\begin{definition}
\label{def:26}
Let $(M,G)$ be a four-dimensional area metric manifold. We call $(M,G)$ a strongly hyperbolic area metric spacetime if
\begin{enumerate}
	\item the matrix $C(k)=C^{\alpha}k_{\alpha}$ is diagonalizable for all spatial covectors $k$ and has only real eigenvalues $\lambda_i$ and
	\item the diagonalisation of $C(k)$ is well-conditioned: if $S(k)C(k)S(k)^{-1}=\mbox{diag}(\lambda_i)$ then $\mbox{sup}_{k\in S^2}=||S(k)^{-1}||~||S(k)||<\infty$.
\end{enumerate}
\end{definition}

The first requirement simply reformulates the weak hyperbolicity requirement of Definition \ref{def:25}, but the second requirement in this definition deserves some further comment. A solution of equation (\ref{eqn:Cauchy}) may be obtained by performing a Fourier transformation on the spatial components to get rid of the spatial derivatives. The resulting ordinary differential equation does only contain time derivatives of the transformed fields $\hat u_N$ and may be solved in standard fashion. To obtain the solution of the original system (\ref{eqn:Cauchy}) we need to perform an inverse Fourier transformation of the fields $\hat u_N$. The second condition in the definition ensures that the $\hat u_N$ are square integrable which means that the inverse Fourier transformation is indeed possible.

It can now be shown that a strongly hyperbolic area metric spacetime renders the initial value problem for Maxwell theory locally well-posed \cite{PDE:1}. This proposition only holds locally since we were investigating the field equations (\ref{eqn:Cauchy}) in a sufficiently small neighborhood of a point $p\in M$ such that the coefficients in (\ref{eqn:Cauchy}) may be assumed constant, but this suffices for our purposes. In the next section, we develop an algebraic classification of area metric manifolds which will be related, in section \ref{chap:4}, to the purely analytic characterisation of area metric spacetimes we gave here in Definitions \ref{def:25} and \ref{def:26}.

Global causality conditions for area metric spacetimes may now be imposed, exactly following the lines known from Lorentzian geometry \cite{Beem}, employing the technical machinery developed above. Recall that future timelike vectors are vectors $X$ that lie in the future timelike vector cone $C_p$ while future causal vectors lie in the topological closure $\overline{C_p}$ of $C_p$. A curve $\gamma:~I\subseteq\mathds{R}\rightarrow M$ is said to be timelike (causal) if its tangent vectors at every point $p\in M$ are timelike (causal). We call an area metric spacetime \textit{chronological} if it does not contain any closed timelike curves. Physically speaking, this means that no one is able to meet himself in the past. The proof of the theorem that any compact area metric spacetime contains closed timelike curves and thus fails to be chronological follows the same lines as the proof in the analogous theorem in Lorentzian geometry. An area metric spacetime $(M,G)$ is called causal if there exist no closed causal curves. This definition is a little more restrictive then $(M,G)$ being chronological. Again physically speaking, this means no one is able to communicate with himself in the past. One may then also define the \textit{chronological future} of a point $p\in M$ as the set $I^+(p)$ of all points $q$ for which there exist a future directed timelike curve $\gamma$ from $p$ to $q$. It may be shown that $I^+(p)$ is an open set. The \textit{causal future} $J^+(p)$ of a point $p$ is defined analogously with the only difference that the curve connecting $p$ and $q$ has to be causal. In contrast to $I^+(p)$, the set $J^+(p)$ is neither open nor closed. Accordingly, the chronological past $I^-(p)$ and the causal past $J^-(p)$ of a point $p$ may be defined.


 With the above notions, one may define even stronger causal requirements for area metric spacetimes by extending the attention to entire neighborhoods $U(p)$ of some point $p$. An area metric spacetime is called \textit{strongly causal} if no causal curve that leaves a sufficiently small neighborhood $U(p)$ of a point $p$ ever returns. Remarkably, one may prove that for a strongly causal area metric spacetime $(M,G)$, the Alexandrov topology (which is generated by all diamonds of the form $I^+(p)\cap I^-(q)$ with $p,q\in M$) agrees with the given topology of the manifold $M$. Again the proof follows the same line of arguments as the proof of the analogous proposition in Lorentzian geometry.

Clearly, we were only able to touch on the very basics of the global causality theory for area metric spacetimes here. As in the case of Lorentzian manifolds, it should be interesting to push this study further, to see which of the other well-known metric theorems directly extend to the area metric case, and what further physical conclusions can be drawn. However, we now temporarily change our focus to an algebraic classification of four-dimensional area metrics in the next section, before we combine the latter with our studies of causality conditions presented above.

\section{Algebraic classification of four-dimensional area metrics}
\label{chap:3}
The algebraic classification of area metrics in four dimensions, which we present in this section, constitutes the second technical pillar on which the results to be derived in section \ref{chap:4} rest. In essence, the here obtained classification amounts to 46 continuous families of distinct algebraic classes of four-dimensional area metrics. Crucially, those may be conveniently grouped into 23 metaclasses, which we explicitly displayed at the end of this section, and which play an important role in deciding which algebraic classes of area metrics can constitute spacetimes.

\subsection{Formulation of the problem}
We first need to decide according to which criterion we want to classify area metrics. Since we are interested in the area metric data that are independent of a choice of frame, we choose to locally identify area metrics that contain the same information up to a change of the local frame. Therefore we will classify area metrics according to the following

\begin{definition}
We call two area metrics $G$ and $H$ on the same $d$-dimensional manifold $M$ strongly equivalent, $G\sim H$, if for every point $p \in M$ there exists a $GL(d)$-transformation $t$ such that
\begin{equation}
G_{abcd}=t_{~~a}^{m}t_{~~b}^{n}t_{~~c}^{p}t_{~~d}^{q}H_{mnpq}\,.
\label{eqn:GHsim}
\end{equation}
\end{definition}

Clearly, the relation $\sim$ is an equivalence relation. The problem of classification can now be stated as follows: identify the algebraic classes of area metrics as the equivalence classes with respect to the equivalence relation $\sim$. In other words, two area metrics $G$ and $G^*$ which cannot be pointwise related by a change of frame, according to (\ref{eqn:GHsim}), belong to different algebraic classes.

Once the equivalence classes are identified, it is convenient to pick a particularly simple representative of each algebraic class, which we will refer to as the normal form of this class.

In two dimensions, area metric geometry is essentially symplectic geometry, as we have seen in section 2. Therefore the classification of area metrics in two dimensions is obtained by virtue of Darboux's theorem for symplectic vectors spaces \cite{Symplectic}. It states that up to frame transformations, there is only one symplectic form. In three dimensions, area metric geometry is essentially metric geometry. Consequently, the classification of area metrics in three dimensions can be carried out with the help of Sylvester's theorem \cite{Lang} for symmetric bilinear forms. From the fact that one needs to employ rather different classification theorems in two and three dimensions, namely Darboux's theorem on the one hand and Sylvester's theorem on the other hand, one may expect that yet another theorem must be found to classify area metrics in four dimensions. In this section, we will show that this is indeed the case. The classification in dimensions higher than four currently remains an open problem.

\subsection{Weak classification of area metrics}
As a first step towards the classification of four-dimensional area metrics with respect to the equivalence relation $\sim$, we will first consider a weak classification of area metrics in arbitrary dimensions $d$. We will then use the insights of these considerations two solve our original classification problem.

From the symmetries of the area metric $G$, it is immediately clear that equation (\ref{eqn:GHsim}) takes the form
\begin{equation}
G_{abcd}=t_{~~a}^{\lbrack m}t_{~~b}^{n\rbrack }t_{~~c}^{\lbrack p}t_{~~d}^{q\rbrack }H_{mnpq}\,.
\label{eqn:ttsym}
\end{equation}
Let us now use Petrov notation to write equation (\ref{eqn:ttsym}) in the form
\begin{equation}
 G_{AB}=T_{~~A}^{M}T_{~~B}^{N} H_{MN},
\label{eqn:coarse}
\end{equation}
where $T_{~~A}^{M}$ is the Petrov matrix associated to the tensor $2t_{~~a}^{\lbrack m}t_{~~b}^{n\rbrack}$,
\begin{equation}
T=Petrov(2t_{~~a}^{\lbrack m}t_{~~b}^{n\rbrack}).
\label{eqn:T}
\end{equation}

Although we introduced $T$ as a transformation induced by $t\in GL(d)$, one may also read (\ref{eqn:coarse}) as an equation for an arbitrary $ T\in GL(d(d-1)/2)$. It is clear that such a $T$ is generally not induced from a $t\in GL(d)$. Hence the requirement that two area metrics be related by a $GL(d(d-1)/2)$ transformation as in (\ref{eqn:coarse}) is weaker then the requirement (\ref{eqn:GHsim}). This gives rise to the following definition:

\begin{definition}
We call two area metrics $G$ and $H$ on the same $d$-dimensional manifold $M$ weakly equivalent, $G\approx H$, if for every point $p \in M$ there exists a $GL(d(d-1)/2)$-transformation $T$ such that equation (\ref{eqn:coarse}) holds.
\end{definition}

Obviously, the relation $\approx$ is also an equivalence relation with respect to which we may classify area metrics. It is also clear that two area metrics $G$ and $G^*$ that are strongly equivalent are automatically weakly equivalent since any $t\in GL(d)$ induces a $T\in GL(d(d-1)/2)$ as we have seen above. However, the converse does not hold. Classification of area metrics with respect to strong equivalence $\sim$ rather amounts to weak classification under the constraint of picking only those $T\in GL(d(d-1)/2)$ that are indeed induced by a $t\in GL(d)$. Finding a condition in four dimensions that ensures that $T$ is of the form (\ref{eqn:T}) is the task solved in the next section.

The classification of $d$-dimensional area metrics with respect to the weak equivalence relation $\approx$ itself can be achieved easily by applying Sylvester's theorem, since the area metric in equation (\ref{eqn:coarse}) defines a symmetric bilinear form on $\mathds{R}^{d(d-1)/2}$. Sylvester's theorem states that the $GL(d(d-1))/2$-signature of such an inner product is the only frame-independent information. We can therefore classify area metrics in $d$ dimensions according to their $GL(d(d-1)/2)$-signature. This amounts to $d(d-1)/2+1$ possible algebraic classes.

\subsection{Strong classification of area metrics in four dimensions}
We may now formulate a condition for a transformation $T\in GL(6)$ to be induced by a $GL(4)$ transformation to refine the weak classification to the algebraic classification we actually aimed for. In four dimensions, there indeed is such a condition, using the fact that the canonical area metric volume form (\ref{eqn:volf}) which in Petrov form reads
\begin{equation}
Petrov(\omega_G)=|\mbox{det}(Petrov(G))|^{1/6}\left[\begin{array}{cccccc}
0&0&0&0&0&1\\
0&0&0&0&1&0\\
0&0&0&1&0&0\\
0&0&1&0&0&0\\
0&1&0&0&0&0\\
1&0&0&0&0&0
\end{array}\right]\, ,
\label{eqn:volfmat}
\end{equation}
is an area metric in its own right. With this in mind, we have the following

\begin{theorem}\label{theo:31}
Let $G$ and $H$ be two area metrics on an orientable four-dimensional manifold $M$. If the weak equivalences $G\approx H$ and $\omega_G\approx \omega_H$ hold simultaneously with the same $GL(6)$ transformation, then we have either the strong equivalence $G\sim H$ or  $G\sim \Sigma^tH\Sigma$, where the components $\Sigma^{ab}_{cd}$ of the endomorphism $\Sigma$ are numerically the same as $\epsilon_{abcd}$ with $\epsilon_{0123}=1$.
\end{theorem}
\textbf{Proof.} With the help of the inverse identification of the capital Petrov indices with antisymmetric pairs of indices $[ab]$ over some given frame $\{e_a\}$ of $\mathds{R}^4$, the weak equivalences may be expressed as
\begin{equation}
G_{abcd}=\frac{1}{4}T_{ab}^{mn}T_{cd}^{pq}H_{mnpq},~~~4|\mbox{det}(Petrov(GH^{-1}))|^{1/6} \epsilon_{abcd}=T_{ab}^{mn}T_{cd}^{pq}\epsilon_{mnpq}\,.
\end{equation}
The second condition can be read as a restriction on the six bivectors $\{ T^{ab}_{01},T^{ab}_{02}$, $T^{ab}_{03},T^{ab}_{12},T^{ab}_{31},T^{ab}_{23}\}$: we must have vanishing $T_{01}\wedge T_{02}$, $T_{02}\wedge T_{03}$ and $T_{03}\wedge T_{01}$ and all $T_{ab}$ must be wedge products of two vectors which is equivalent to have vanishing $T_{ab}\wedge T_{ab}$ (no sum). The first three conditions can be solved in two inequivalent ways: either $T_{01}$, $T_{02}$ and $T_{03}$ have a direction in common, or they pairwise intersect. But this precisely corresponds to either $T_{cd}^{ab}=t_{~~c}^{\lbrack a}t^{b\rbrack}_{~~d}$ or $T_{cd}^{ab}=t_{~~c}^{\lbrack m}t^{n\rbrack}_{~~d}\Sigma_{mn}^{ab}$ in terms of some $GL(4)$ transformation with $\mbox{det}(t)>0$ \cite{Reisenberger:1995xh}, \cite{DePietri:1998mb}. Since $M$ is orientable we can consistently restrict our attention to this case, which concludes the proof.\\

The result of this theorem deserves some further comments.
The correspondence between the weak and the strong equivalence involves an ambiguity in the order of frame on $\mathds{R}^6$ by means of the $\Sigma$-symbols. This only implies that once we have found suitable normal forms of the simultaneous weak classification of $G$ and $\omega_G$, we must be careful with the interpretation of the Petrov matrix representing the normal forms. We will return to that point later.

Fortunately, the remaining step in obtaining the desired strong classification of the area metric $G$ now reduces to the problem of the simultaneous weak classification of the area metric $G$ and its associated area metric volume form $\omega_G$. The solution to this problem is known, and we cite the relevant theorem without proof \cite{Algebra:1} in a form that is congenial for our purpose.

\begin{theorem}\label{theo:32}
Let $G$ and $\omega_G$ be two symmetric bilinear forms on $\mathds{R}^6$. Then there exists a basis on $\mathds{R}^6$ such that the matrices that represent $G$ and $\omega_G$ take the following block diagonal form:
\begin{alignat*}{6}
Petrov(G) =&R_1  &\oplus  \ldots \oplus &~ R_{m}  &\oplus  C_{m+1} \oplus \ldots \oplus &  C_n,\\
Petrov(\omega_G) =& \epsilon_1 P_1 &\oplus   \ldots  \oplus &   \epsilon_mP_m &\oplus  P_{m+1} \oplus \ldots \oplus & P_{n},
\end{alignat*}
where blocks with the same index have equal size and the matrices representing the blocks $R_p$ are of the form
\begin{equation*}
R_p(\lambda_p)=\left[\begin{array}{cccc}
0&\cdots &0 & \lambda_p   \\
\vdots& & \iddots& 1   \\
0&\iddots & \iddots& 0   \\
\lambda_p&1 & 0&  0  \\
\end{array}\right]
\end{equation*}
with real numbers $\lambda_p$ (which incidentally correspond to the (real) eigenvalues of the endomorphism $J=\omega_G^{-1}G$), whereas the $C_q$  take the form
\begin{equation*}
C_q(\sigma_q,\tau_q)=\left[\begin{array}{cccccc}
0& 0&0& 0& -\tau_q&\sigma_q \\
0&0 &0 &0 &\sigma_q &\tau_q \\
0&0 && \iddots& 0&1 \\
0&0 & \iddots & & 1&0 \\
 -\tau_q&\sigma_q& 0&1 &0 &0 \\
\sigma_q &\tau_q&1 &0 &0 &0
\end{array}\right].
\end{equation*}
with real numbers $\sigma_q$ and $\tau_q$ (corresponding to the (complex) eigenvalues $\sigma\pm i \tau$ of $J$) with $\tau_q >0$. Finally, we have the signs $\epsilon_j=\pm 1$ and
\begin{equation*}
P_j=\left[\begin{array}{cccc}
0& \cdots&0 & 1\\
\vdots& &\iddots &0 \\
0&\iddots & &\vdots   \\
1&0 &\cdots &0  \\
\end{array}\right].
\end{equation*}
\end{theorem}

This theorem now applies to our problem as follows. Given an area metric $G$ and its associated volume form $\omega_G$ in Petrov notation we use the theorem to bring $G$ and $\omega_G$ to the stated simultaneous normal forms by a $GL(6)$ transformation. After that we still need to apply a further change of basis to bring $\omega_G$ to the form of equation (\ref{eqn:volfmat}) so that we can apply our Theorem \ref{theo:31}. Such a change of basis on $\mathds{R}^6$ is always possible if the matrix provided by the theorem that represents $\omega_G$ has $GL(6)$ signature $(3,3)$, since we know that the signature is the only local frame-independent information for such a symmetric bilinear form. The change of basis has to be simultaneously applied to the area metric $G$, and the resulting matrix then is the desired normal form of the area metric.  Thus all $GL(6)$-inequivalent pairs $(G,\omega_G)$ appearing in Theorem \ref{theo:32} for which $\omega_G$ has $GL(6)$-signature $(3,3)$ represent a different algebraic class of the area metric $G$. Distinguishing the different sign characteristics of the blocks in $\omega_G$, a simple counting reveals that there are 46 distinct algebraic classes.\\
There is another subtlety in the application of the theorem. To make sure that the obtained pair of normal forms $(G,\omega_G)$ is a pair of an area metric and its associated volume form we need to require $|\mbox{det}(Petrov(G))|=-\mbox{det}(Petrov((\omega_G))=1$ which follows directly from the definition of the volume form. This requirement constraints the scalars in the normal form of the area metric $G$. The theorem itself determines up to six scalars and the extra condition on $\mbox{det}(Petrov(G))$ renders only five of them independent. This confirms our intuitive claim from section 2 where we suspected a four-dimensional area metric to determine up to five $GL(4)$-scalars.\\
Before we continue, we illustrate the above-described procedure for obtaining the normal form of a given area metric $G$.

\textbf{Example.} Consider the following possible pair of matrices provided by Theorem \ref{theo:32} that represent two bilinear forms $G$ and $\omega_G$
\begin{equation}
Petrov(\omega_G)=\left[\begin{array}{cccccc}
0&1&0&0&0&0\\
1&0&0&0&0&0\\
0&0&0&1&0&0\\
0&0&1&0&0&0\\
0&0&0&0&0&1\\
0&0&0&0&1&0
\end{array}\right]\, ,
\label{eqn:twomatrices1}
\end{equation}
\begin{equation}
Petrov(G)=\left[\begin{array}{cccccc}
-\tau_1&\sigma_1 &0 &0 & 0&0 \\
\sigma_1&\tau_1 &0&0 &0 &0  \\
0&0 & -\tau_2&\sigma_2 &0 &0\\
0&0 & \sigma_2&\tau_2 &0 &0\\
0 &0 &0 &0&-\tau_3&\sigma_3\\
0 &0 &0 &0&\sigma_3&\tau_3\\
\end{array}\right]\,.
\label{eqn:twomatrices2}
\end{equation}
To recover $\omega_G$ in the form (\ref{eqn:volfmat}) (which is possible in the first place since $Petrov(\omega_G)$ indeed has signature $(3,3)$), we interchange the first and fifth basis vector on $\mathds{R}^6$ which amounts to a change of the first and fifth row and column in $Petrov(\omega_G)$ and $Petrov(G)$. Then we also exchange the second and fifth basis vector. The matrix representing $G$ then takes the normal form
\begin{equation}
Petrov(G)=\left[\begin{array}{cccccc}
-\tau_1&0&0&0&0&\sigma_1\\
0&-\tau_3&0&0&\sigma_3&0\\
0&0&-\tau_2&\sigma_2&0&0\\
0&0&\sigma_2&\tau_2&0&0\\
0&\sigma_3&0&0&\tau_3&0\\
\sigma_1&0&0&0&0&\tau_1
\end{array}\right].
\label{eqn:classI}
\end{equation}
We have to keep in mind that the six scalars in $Petrov(G)$ have to satisfy the condition $|\mbox{det}(Petrov((G))|=1$.\\

We should emphasize an important point. The scalars in $Petrov(G)$ locally determine the area metric completely. Actually every different set of scalars in any of the 46 distinct algebraic classes provided by Theorem \ref{theo:32} determines a separate algebraic class for area metrics. In other words, there are infinitely many algebraic classes of four-dimensional area metrics.

Having reduced the Petrov matrix $Petrov(G)$ of a given area metric $G$ to its normal form according to Theorem \ref{theo:32} we can apply Theorem \ref{theo:31} to find the actual normal forms of the area metric $G$ with respect to $\sim$. We only need to be careful with the mentioned ambiguity by means of the $\Sigma$-symbols in Theorem \ref{theo:31} when we identify the entries of $Petrov(G)$ with components $G_{abcd}$ of the area metric $G$. Both $G_{abcd}$ and $G_{mnpq}\Sigma_{ab}^{mn}\Sigma_{cd}^{pq}$ are the components of distinct normal forms of the area metric $G$.

It proves useful to group the infinitely many normal forms in four dimensions into coarser classes labeled by the Segr\'e type of the endomorphism $J$ appearing in Theorem \ref{theo:32}. We introduce these metaclasses in the following section and then display the resulting families of normal forms.

\subsection{Metaclasses and normal forms}
\label{sec:34}
A convenient way to group the possible normal forms of the area metric $G$ is a division into metaclasses labeled by the Segr\'e type \cite{Segre} of the endomorphism $J$ defined in Theorem \ref{theo:32}. The Segr\'e types of the endomorphism $J$ only take into account the size of the Jordan blocks \cite{Lang} in $J$, and whether the eigenvalues of the corresponding block are complex or real. That is, a Segr\'e type is given by a symbol $[A\bar A~\dots BCD]$ where $A,B,C,D$ are positive integers. If an integer $A$ in the label is followed by $\bar A$, the endomorphism $J$ contains a Jordan block of size $A$ with a complex eigenvalue and a block with the same size and the complex conjugate eigenvalue. Otherwise the endomorphism contains a real Jordan Block of size $B$, $C$ and $D$. For example, the normal form $Petrov(G)$ in (\ref{eqn:classI}) is of Segr\'e type $[1\bar 1~1\bar 1~1\bar 1]$ because the corresponding endomorphism $J$ has six distinct complex eigenvalues where three of them are simply the complex conjugates of the other three, and the Jordan block for every eigenvalue has size one.

The metaclasses of area metrics, labeled as defined by the various Segr\'e types, disregard both the signs $\epsilon_j$ as they appear in Theorem \ref{theo:32}, and the actual eigenvalues of the endomorphism $J$. This gives rise to $23$ different metaclasses of area metrics:
\begin{itemize}
\item three metaclasses where the Jordan blocks of the corresponding endomorphism $J$ only have complex eigenvalues $\sigma_i\pm i\tau_i$
\begin{equation*}
I=[1\bar 1~1\bar 1~1\bar 1],~~~II=[1\bar 1~2\bar 2],~~~III=[3\bar 3],
\end{equation*}
\item four metaclasses with real Jordan blocks in $J$ of at most size one
\begin{equation*}
IV=[1\bar 1~1\bar 1~11],~~V=[2\bar 2~11],~~VI=[1\bar 1~1111],~~VII=[111111]
\end{equation*}
\item 16 metaclasses with at least one real Jordan block in $J$ of size greater or equal to two.
\end{itemize}

These metaclasses prove very useful. Indeed, in the next section we will present a powerful theorem that renders the 16 metaclasses VIII-XXIII (which feature real Jordan blocks of size greater or equal to two) as unphysical since these do not define strongly hyperbolic area metric spacetimes as defined in the previous section.

Now that we have introduced the metaclasses of area metrics labeled by the Segr\'e types of the endomorphism $J$, we may present a complete list of normal forms of these metaclasses.

\begin{theorem}\label{theo:33}
Let $(M,G)$ be a four-dimensional area metric manifold. Then at each point $p\in M$there exists a frame $\{e_a\}$ in which the Petrov matrix $Petrov(G)$ of $G$ takes one of the following forms.

\vspace{-0.5cm}
\begin{center}
\centering
 \begin{minipage}[t]{50mm}
 \centering
 \vspace{0.5cm}
 metaclass I $[1\bar{1}~1\bar{1}~1\bar{1}]$
 \begin{equation*}
 \left[\begin{array}{cccccc}
-\tau_1&0&0&0&0&\sigma_1\\
0&-\tau_3&0&0&\sigma_3&0\\
0&0&-\tau_2&\sigma_2&0&0\\
0&0&\sigma_2&\tau_2&0&0\\
0&\sigma_3&0&0&\tau_3&0\\
\sigma_1&0&0&0&0&\tau_1
\end{array}\right]
 \end{equation*}
 \end{minipage}
 \hspace{20mm}
\begin{minipage}[t]{50mm}
 \centering
 \vspace{0.5cm}
 metaclass II $[2\bar{2}~1\bar{1}]$
 \begin{equation*}
 \left[\begin{array}{cccccc}
0&0&0&0&-\tau_1&\sigma_1\\
0&0&0&0&\sigma_1&\tau_1\\
0&0&-\tau_2&\sigma_2&0&0\\
0&0&\sigma_2&\tau_2&0&0\\
-\tau_1&\sigma_1&0&0&0&1\\
\sigma_1&\tau_1&0&0&1&0
\end{array}\right]
 \end{equation*}
 \end{minipage}

\centering
\begin{minipage}[t]{50mm}
 \centering
 \vspace{0.5cm}
 metaclass III $[3\bar{3}]$
 \begin{equation*}
 \left[\begin{array}{cccccc}
0&0&0&0&-\tau_1&\sigma_1\\
0&0&0&0&\sigma_1&\tau_1\\
0&0&-\tau_1&\sigma_1&0&1\\
0&0&\sigma_1&\tau_1&1&0\\
-\tau_1&\sigma_1&0&1&0&0\\
\sigma_1&\tau_1&1&0&0&0
\end{array}\right]
 \end{equation*}
 \end{minipage}

\centering
 \begin{minipage}[t]{50mm}
 \centering
 \vspace{0.5cm}
 metaclass IV $[1\bar{1}~1\bar{1}~11]$
 \begin{equation*}
 \left[\begin{array}{cccccc}
-\tau_1&0&0&0&0&\sigma_1\\
0&-\tau_2&0&0&\sigma_2&0\\
0&0&\lambda_1&\lambda_2&0&0\\
0&0&\lambda_2&\lambda_1&0&0\\
0&\sigma_2&0&0&\tau_2&0\\
\sigma_1&0&0&0&0&\tau_1
\end{array}\right]
 \end{equation*}
 \end{minipage}
 \hspace{20mm}
\begin{minipage}[t]{50mm}
 \centering
 \vspace{0.5cm}
 metaclass V $[2\bar{2}~11]$
 \begin{equation*}
 \left[\begin{array}{cccccc}
0&0&0&0&-\tau_1&\sigma_1\\
0&0&0&0&\sigma_1&\tau_1\\
0&0&\lambda_1&\lambda_2&0&0\\
0&0&\lambda_2&\lambda_1&0&0\\
-\tau_1&\sigma_1&0&0&0&1\\
\sigma_1&\tau_1&0&0&1&0
\end{array}\right]
 \end{equation*}
 \end{minipage}

\centering
 \begin{minipage}[t]{50mm}
 \centering
 \vspace{0.5cm}
 metaclass VI $[1\bar{1}~11~11]$
 \begin{equation*}
 \left[\begin{array}{cccccc}
-\tau_1&0&0&0&0&\sigma_1\\
0&\lambda_3&0&0&\lambda_4&0\\
0&0&\lambda_1&\lambda_2&0&0\\
0&0&\lambda_2&\lambda_1&0&0\\
0&\lambda_4&0&0&\lambda_3&0\\
\sigma_1&0&0&0&0&\tau_1
\end{array}\right]
 \end{equation*}
 \end{minipage}
 \hspace{20mm}
\begin{minipage}[t]{50mm}
 \centering
 \vspace{0.5cm}
 metaclass VII $[11~11~11]$
 \begin{equation*}
\left[\begin{array}{cccccc}
\lambda_5&0&0&0&0&\lambda_6\\
0&\lambda_3&0&0&\lambda_4&0\\
0&0&\lambda_1&\lambda_2&0&0\\
0&0&\lambda_2&\lambda_1&0&0\\
0&\lambda_4&0&0&\lambda_3&0\\
\lambda_6&0&0&0&0&\lambda_5
\end{array}\right]
 \end{equation*}
 \end{minipage}
\end{center}

\vspace{0.5cm}
The remaining 16 metaclasses involve a choice of signs $\epsilon_i$ that take the values $\pm 1$. Any combination of these signs denotes a different algebraic class.
\begin{center}
\begin{minipage}[t]{50mm}
 \centering
 \vspace{0.5cm}
 metaclass VIII $[6]$
 \begin{equation*}
 \left[\begin{array}{cccccc}
0&0&0&0&0&\lambda_1\\
0&0&0&0&\lambda_1&1\\
0&0&0&\lambda_1&1&0\\
0&0&\lambda_1&\epsilon_1&0&0\\
0&\lambda_1&1&0&0&0\\
\lambda_1&1&0&0&0&0
\end{array}\right] \end{equation*}
 \end{minipage}
 \hspace{20mm}
\begin{minipage}[t]{50mm}
 \centering
 \vspace{0.5cm}
 metaclass IX $[42]$
 \begin{equation*}
  \left[\begin{array}{cccccc}
0&0&0&0&0&\lambda_1\\
0&0&0&0&\lambda_1&1\\
0&0&0&\lambda_2&0&0\\
0&0&\lambda_2&\epsilon_1&0&0\\
0&\lambda_1&0&0&\epsilon_2&0\\
\lambda_1&1&0&0&0&0
\end{array}\right]
\end{equation*}
 \end{minipage}
\vspace{0.5cm}

\begin{minipage}[t]{50mm}
 \centering
 \vspace{0.5cm}
 metaclass X $[4~1\bar{1}]$
 \begin{equation*}
  \left[\begin{array}{cccccc}
0&0&0&0&0&\lambda_1\\
0&0&0&0&\lambda_1&1\\
0&0&-\tau_1&\sigma_1&0&0\\
0&0&\sigma_1&\tau_1&0&0\\
0&\lambda_1&0&0&\epsilon_1&0\\
\lambda_1&1&0&0&0&0
\end{array}\right]
\end{equation*}
 \end{minipage}
 \hspace{20mm}
\begin{minipage}[t]{50mm}
 \centering
 \vspace{0.5cm}
 metaclass XI $[411]$
 \begin{equation*}
  \left[\begin{array}{cccccc}
0&0&0&0&0&\lambda_1\\
0&0&0&0&\lambda_1&1\\
0&0&\lambda_3&\lambda_2&0&0\\
0&0&\lambda_2&\lambda_3&0&0\\
0&\lambda_1&0&0&\epsilon_1&0\\
\lambda_1&1&0&0&0&0
\end{array}\right]
\end{equation*}
 \end{minipage}
\vspace{0.5cm}

\begin{minipage}[t]{50mm}
 \centering
 \vspace{0.5cm}
 metaclass XII $[2\bar{2}~2]$
 \begin{equation*}
  \left[\begin{array}{cccccc}
0&0&0&0&\tau_1&\sigma_1\\
0&0&0&0&\sigma_1&\tau_1\\
0&0&0&\lambda_1&0&0\\
0&0&\lambda_1&\epsilon_1&0&0\\
-\tau_1&\sigma_1&0&0&0&1\\
\sigma_1&\tau_1&0&0&1&0
\end{array}\right]
\end{equation*}
 \end{minipage}
 \hspace{20mm}
\begin{minipage}[t]{50mm}
 \centering
 \vspace{0.5cm}
 metaclass XIII $[222]$
 \begin{equation*}
  \left[\begin{array}{cccccc}
0&0&0&0&0&\lambda_1\\
0&\epsilon_2&0&0&\lambda_2&0\\
0&0&0&\lambda_3&0&0\\
0&0&\lambda_3&\epsilon_3&0&0\\
0&\lambda_2&0&0&0&0\\
\lambda_1&0&0&0&0&\epsilon_1
\end{array}\right]
\end{equation*}
 \end{minipage}
\vspace{0.5cm}

\begin{minipage}[t]{50mm}
 \centering
 \vspace{0.5cm}
 metaclass XIV $[22~1\bar{1}]$
 \begin{equation*}
  \left[\begin{array}{cccccc}
0&0&0&0&0&\lambda_1\\
0&\epsilon_2&0&0&\lambda_2&0\\
0&0&-\tau_1&\sigma_1&0&0\\
0&0&\sigma_1&\tau_1&0&0\\
0&\lambda_2&0&0&0&0\\
\lambda_1&0&0&0&0&\epsilon_1
\end{array}\right]
\end{equation*}
 \end{minipage}
 \hspace{20mm}
\begin{minipage}[t]{50mm}
 \centering
 \vspace{0.5cm}
 metaclass XV $[2211]$
 \begin{equation*}
  \left[\begin{array}{cccccc}
0&0&0&0&0&\lambda_1\\
0&\epsilon_2&0&0&\lambda_2&0\\
0&0&\lambda_3&\lambda_4&0&0\\
0&0&\lambda_4&\lambda_3&0&0\\
0&\lambda_2&0&0&0&0\\
\lambda_1&0&0&0&0&\epsilon_1
\end{array}\right]\end{equation*}
 \end{minipage}
\vspace{0.5cm}

\begin{minipage}[t]{50mm}
 \centering
 \vspace{0.5cm}
 metaclass XVI $[2~1\bar{1}~1\bar{1}]$
 \begin{equation*}
  \left[\begin{array}{cccccc}
0&0&0&0&0&\lambda_1\\
0&-\tau_2&0&0&\sigma_2&0\\
0&0&-\tau_1&\sigma_1&0&0\\
0&0&\sigma_1&\tau_1&0&0\\
0&\sigma_2&0&0&\tau_2&0\\
\lambda_1&0&0&0&0&\epsilon_1
\end{array}\right]
\end{equation*}
 \end{minipage}
 \hspace{20mm}
\begin{minipage}[t]{50mm}
 \centering
 \vspace{0.5cm}
 metaclass XVII $[211~1\bar{1}]$
 \begin{equation*}
  \left[\begin{array}{cccccc}
0&0&0&0&0&\lambda_1\\
0&-\tau_2&0&0&\sigma_2&0\\
0&0&\lambda_2&\lambda_3&0&0\\
0&0&\lambda_3&\lambda_2&0&0\\
0&\sigma_2&0&0&\tau_2&0\\
\lambda_1&0&0&0&0&\epsilon_1
\end{array}\right]
\end{equation*}
 \end{minipage}
\vspace{0.5cm}

\begin{minipage}[t]{50mm}
 \centering
 \vspace{0.5cm}
 metaclass XVIII $[21111]$
 \begin{equation*}
  \left[\begin{array}{cccccc}
0&0&0&0&0&\lambda_1\\
0&\lambda_4&0&0&\lambda_5&0\\
0&0&\lambda_2&\lambda_3&0&0\\
0&0&\lambda_3&\lambda_2&0&0\\
0&\lambda_5&0&0&\lambda_4&0\\
\lambda_1&0&0&0&0&\epsilon_1
\end{array}\right] \end{equation*}
 \end{minipage}
\vspace{0.5cm}

\begin{minipage}[t]{80mm}
\centering
 \vspace{0.5cm}
 metaclass XIX $[51]$
 \begin{equation*}
 \left[\begin{array}{cccccc}
0&0&0&0&0&\lambda_1\\
0&0&0&0&\lambda_1&1\\
0&0&\frac{\epsilon_1}{2}(\lambda_1-\lambda_2)&\frac{1}{2}(\lambda_1+\lambda_2)&\frac{1}{\sqrt{2}}&0\\
0&0&\frac{1}{2}(\lambda_1+\lambda_2)&\frac{\epsilon_1}{2}(\lambda_1-\lambda_2)&\frac{\epsilon_1}{\sqrt{2}}&0\\
0&\lambda_1&\frac{1}{\sqrt{2}}&\frac{\epsilon_1}{\sqrt{2}}&0&0\\
\lambda_1&1&0&0&0&0
\end{array}\right]
\end{equation*}
\end{minipage}
\vspace{0.5cm}

\begin{minipage}[t]{80mm}
 \centering
 \vspace{0.5cm}
 metaclass XX $[33]$
 \begin{equation*}
 \left[\begin{array}{cccccc}
0&0&0&0&0&\lambda_1\\
0&\frac{1}{2}(\lambda_1-\lambda_2)&0&-\frac{1}{\sqrt{2}}&\frac{1}{2}(\lambda_1+\lambda_2)&\frac{1}{\sqrt{2}}\\
0&0&0&\lambda_2&0&0\\
0&-\frac{1}{\sqrt{2}}&\lambda_2&0&\frac{1}{\sqrt{2}}&0\\
0&\frac{1}{2}(\lambda_1+\lambda_2)&0&\frac{1}{\sqrt{2}}&\frac{1}{2}(\lambda_1-\lambda_2)&\frac{1}{\sqrt{2}}\\
\lambda_1&\frac{1}{\sqrt{2}}&0&0&\frac{1}{\sqrt{2}}&0
\end{array}\right]
\end{equation*}
 \end{minipage}
\vspace{0.5cm}

\begin{minipage}[t]{80mm}
 \centering
 \vspace{0.5cm}
 metaclass XXI $[321]$
 \begin{equation*}
 \left[\begin{array}{cccccc}
\epsilon_2&0&0&0&0&\lambda_2\\
0&0&\frac{\epsilon_1}{\sqrt{2}}&\frac{1}{\sqrt{2}}&\lambda_1&0\\
0&\frac{\epsilon_1}{\sqrt{2}}&\frac{\epsilon_1}{2}(\lambda_1-\lambda_3)&\frac{1}{2}(\lambda_1+\lambda_3)&0&0\\
0&\frac{1}{\sqrt{2}}&\frac{1}{2}(\lambda_1+\lambda_3)&\frac{\epsilon_1}{2}(\lambda_1-\lambda_3)&0&0\\
0&\lambda_1&0&0&0&0\\
\lambda_2&0&0&0&0&0
\end{array}\right]\end{equation*}
 \end{minipage}
\vspace{0.5cm}

\begin{minipage}[t]{80mm}
 \centering
 \vspace{0.5cm}
 metaclass XXII $[31~1\bar{1}]$
 \begin{equation*}
 \left[\begin{array}{cccccc}
-\tau_1&0&0&0&0&\sigma_1\\
0&0&\frac{\epsilon_1}{\sqrt{2}}&\frac{1}{\sqrt{2}}&\lambda_1&0\\
0&\frac{\epsilon_1}{\sqrt{2}}&\frac{\epsilon_1}{2}(\lambda_1-\lambda_2)&\frac{1}{2}(\lambda_1+\lambda_2)&0&0\\
0&\frac{1}{\sqrt{2}}&\frac{1}{2}(\lambda_1+\lambda_2)&\frac{\epsilon_1}{2}(\lambda_1-\lambda_2)&0&0\\
0&\lambda_1&0&0&0&0\\
\sigma_1&0&0&0&0&\tau_1
\end{array}\right]\end{equation*}
 \end{minipage}
\vspace{0.5cm}

\begin{minipage}[t]{80mm}
 \centering
 \vspace{0.5cm}
 metaclass XXIII $[3111]$
 \begin{equation*}
 \left[\begin{array}{cccccc}
\lambda_3&0&0&0&0&\lambda_4\\
0&\frac{\epsilon_1}{2}(\lambda_1-\lambda_2)&\frac{\epsilon_1}{\sqrt{2}}&0&\frac{1}{2}(\lambda_1+\lambda_2)&0\\
0&\frac{\epsilon_1}{\sqrt{2}}&0&\lambda_1&\frac{1}{\sqrt{2}}&0\\
0&0&\lambda_1&0&0&0\\
0&\frac{1}{2}(\lambda_1+\lambda_2)&\frac{1}{\sqrt{2}}&0&\frac{\epsilon_1}{2}(\lambda_1-\lambda_2)&0\\
\lambda_4&0&0&0&0&\lambda_3
\end{array}\right]\end{equation*}
 \end{minipage}

\end{center}
\end{theorem}

With the above theorem guaranteeing the existence of a frame such that an area metric takes one of the listed 23 normal forms, it only remains to study whether such a frame is uniquely determined or, if not, how it is related to other such frames. This question is addressed by the following theorem, already anticipating the result from the next section that the metaclasses VIII to XXIII must be discarded as physically not viable area metric spacetimes, see section 4.1.
\newpage
\begin{theorem}
Let $(M,G)$ be a four-dimensional area metric manifold with an area metric falling into one of the metaclasses I to VII. Then the frame in which the Petrov matrix $Petrov(G)$ representing $G$ takes the form displayed in theorem \ref{theo:33} is determined up to a transformation obtained by exponentiation of the algebras presented in table \ref{tab:symtab}.
\begin{table*}[h!]
	\centering
		\begin{tabular}{|c|c|c|}
		\hline
    metaclass &local gauge algebra... & ...in presence of degeneracies\\
    \hline
    I     &  $o(1,1)\oplus o(2)$ & w.l.o.g. $\sigma_1=\sigma_2$, $\tau_1=\tau_2$\\
          &  $o(1,3)$           &all $\sigma_i=\sigma_j$, $\tau_i=\tau_j$\\
    \hline
    II    &     see generators (\ref{eqn:classII})& $\tau_1=\tau_2$ and $\sigma_1=\sigma_2$     \\
    \hline
    III   &        no symmetries          & \\
    \hline
    IV    &  $o(2)$             & $\tau_i=\lambda_1$ and $\sigma_i=\lambda_2$, $i=1\vee i=2$\\
          &  $o(1,1)\oplus o(2)$& $\tau_1=\tau_2$ and $\sigma_1=\sigma_2$\\
          &  $o(1,1)\oplus o(2)\oplus o(2)\oplus o(2)$& $\tau_1=\tau_2=\lambda_1$ and $\sigma_1=\sigma_2=\lambda_2$\\
    \hline
    V    &         no symmetries             &\\
    \hline
    VI   &        $o(2)$        & $\tau_1=\lambda_1$ and $\sigma_1=\lambda_2$ or $\tau_1=\lambda_3$ and $\sigma_1=\lambda_4$\\
         &  $o(2)\oplus o(2)$   & $\lambda_1=\lambda_3$ and $\lambda_2=\lambda_4$\\
         &  $o(2)\oplus o(2)\oplus o(2)\oplus o(2)$& $\tau_1=\lambda_1=\lambda_3$ and $\sigma_1=\lambda_2=\lambda_4$\\
    \hline
    VII  &   $o(2)\oplus o(2)$   & w.l.o.g. $\lambda_1=\lambda_3$ and $\lambda_2=\lambda_4$\\
         &   $o(4)$              & $\lambda_1=\lambda_3=\lambda_5$ and $\lambda_2=\lambda_4=\lambda_6$\\
    \hline    			
		\end{tabular}
		\caption{Local gauge algebras for metaclass I-VII area metrics}
		\label{tab:symtab}
\end{table*}
The elements of the possible gauge algebra of metaclass II area metrics are of the form
\begin{equation}
\omega^{II}_1=\left(\begin{array}{cccc}
0&0&0&1\\
0&0&0&0\\
0&0&0&0\\
0&1&0&0
\end{array}\right),~~~
\omega^{II}_2=\left(\begin{array}{cccc}
0&0&1&0\\
0&0&0&0\\
0&1&0&0\\
0&0&0&0
\end{array}\right)\, .
\label{eqn:classII}
\end{equation}
\end{theorem}
\textbf{Proof.} The requirement that the Petrov matrix $Petrov(G)$ representing $G$ stay invariant under a change of the local frame can be expressed in a given basis $\{e_a\}$ as
\begin{equation}
t_{~~a}^{m}t_{~~b}^{n}t_{~~c}^{p}t_{~~d}^{q}G_{mnpq}=G_{abcd}.
\label{eqn:bedingung}
\end{equation}
Without loss of generality, we may assume that $G$ is given in normal form, and anticipating the unphysicality of metaclasses VIII-XXIII shown in section \ref{sec:41},we restrict attention to metaclasses I to VII. Focusing on the connected component of the identity of the invariance group, we consider infinitesimal transformations of the form $t_{~~b}^a=\delta_b^a+h\omega_{~~b}^a$ with infinitesimally small $h$ and generators $\omega_{~~b}^a$. Equation (\ref{eqn:bedingung}) then reads
\begin{equation}
0=\omega_{~~a}^mG_{mbcd}+\omega_{~~b}^mG_{amcd}+\omega_{~~c}^mG_{abmd}+\omega_{~~d}^mG_{abcm}\,.
\label{eqn:computesym}
\end{equation}
These are 21 equations for the sixteen components $\omega_{~~b}^a$ of the generator. This system can now be analyzed for the various metaclasses I to VII. We illustrate the procedure of calculating the generators $\omega$ for metaclass I area metrics (\ref{eqn:classI}).\\
For a metaclass I area metric $G$ in normal form (\ref{eqn:classI}) the 21 equations (\ref{eqn:computesym}) decouple into two sets of equations, nine for the diagonal elements of the generators $\omega$, and 12 for the off-diagonal elements. The nine coupled homogeneous equations for the diagonal elements $\omega_{~~i}^i$ are
\begin{eqnarray*}
\sigma_1(\omega_{~~0}^0+\omega_{~~1}^1+\omega_{~~2}^2+\omega_{~~3}^3)&=&0,\\
\sigma_2(\omega_{~~0}^0+\omega_{~~1}^1+\omega_{~~2}^2+\omega_{~~3}^3)&=&0,\\
\sigma_3(\omega_{~~0}^0+\omega_{~~1}^1+\omega_{~~2}^2+\omega_{~~3}^3)&=&0,
\end{eqnarray*}
\vspace{-15mm}
\begin{alignat*}{2}
\tau_1(\omega_{~~0}^0+\omega_{~~1}^1)=0,~~&~~
\tau_1(\omega_{~~2}^2+\omega_{~~3}^3)=0,\\
\tau_2(\omega_{~~0}^0+\omega_{~~3}^3)=0,~~&~~
\tau_2(\omega_{~~1}^1+\omega_{~~2}^2)=0,\\
\tau_3(\omega_{~~0}^0+\omega_{~~2}^2)=0,~~&~~
\tau_3(\omega_{~~1}^1+\omega_{~~3}^3)=0.
\end{alignat*}
If the area metric is non-degenerate, i.e. $\sigma_n\not= 0$ and $\tau_n\not= 0$, only four of these nine equations are independent, and then imply that all diagonal elements vanish.

Further there are 12 coupled homogeneous equations for the 12 unknown off-diagonal elements $\omega_{~~b}^a$. We may write these equations as a matrix equation $A\cdot x=0$ for the vector
\begin{equation*} x=(\omega_{~~1}^0,\omega_{~~0}^1,\omega_{~~2}^0,\omega_{~~0}^2,\omega_{~~3}^0,\omega_{~~0}^3,\omega_{~~2}^1,\omega_{~~1}^2,\omega_{~~3}^1,\omega_{~~1}^3,\omega_{~~3}^2,\omega_{~~2}^3)
\end{equation*}
and the matrix
\begin{equation*}
A=\left[\begin{array}{cccccccccccc}
\tau_2&-\tau_3&&&&&&&&&\sigma_{32}& \\
\tau_3&-\tau_2&&&&&&&&&&\sigma_{32}\\
&&\tau_2&-\tau_1&&&&&\sigma_{21}&&&\\
&&\tau_1&-\tau_2&&&&&&\sigma_{21}&&\\
&&&&\tau_3&-\tau_1&\sigma_{13}&&&&&\\
&&&&\tau_1&-\tau_3&&\sigma_{13}&&&&\\
&&&&\sigma_{31}&&\tau_3&\tau_1&&&&\\
&&&&&\sigma_{13}&\tau_1&\tau_3&&&&\\
&&\sigma_{12}&&&&&&\tau_2&\tau_1&&\\
&&&\sigma_{21}&&&&&\tau_1&\tau_2&&\\
\sigma_{23}&&&&&&&&&&\tau_2&\tau_3\\
&\sigma_{32}&&&&&&&&&\tau_3&\tau_2
\end{array}\right],
\end{equation*}
where we used the shorthand $\sigma_{ij}=\sigma_i-\sigma_j$. The symmetry generators are obviously the non-trivial solutions of the system $A\cdot x=0$. Such solutions do only exist if $\mbox{det}(A)=0$.
Now we observe that if for all pairs $(\sigma_i,\tau_i)\not=(\sigma_j,\tau_j)$ for $i\not=j$ the matrix $A$ has full rank and thus $\omega^i_{~~j}=0$ for all $i,j$. In this case the area metric has no gauge symmetries at all.\\
Let us now have $\sigma_1=\sigma_2$ and $\tau_1=\tau_2$. Then we have non-trivial solutions for $\omega^2_{~~0}$, $\omega^0_{~~2}$, $\omega^3_{~~1}$ and $\omega^1_{~~3}$. We may write the two resulting generators $\omega_1$ and $\omega_2$ in matrix form,
\begin{equation}
\omega_1=\left(\begin{array}{cccc}
0&0&1&0\\
0&0&0&0\\
1&0&0&0\\
0&0&0&0
\end{array}\right),~~~
\omega_2=\left(\begin{array}{cccc}
0&0&0&0\\
0&0&0&1\\
0&0&0&0\\
0&-1&0&0
\end{array}\right)\,.
\end{equation}
In like fashion we find the generators $\omega_3$ and $\omega_4$ if $\sigma_1=\sigma_3$ and $\tau_1=\tau_3$:
\begin{equation}
\omega_3=\left(\begin{array}{cccc}
0&0&0&1\\
0&0&0&0\\
0&0&0&0\\
1&0&0&0
\end{array}\right),~~~
\omega_4=\left(\begin{array}{cccc}
0&0&0&0\\
0&0&-1&0\\
0&1&0&0\\
0&0&0&0
\end{array}\right)\, .
\end{equation}
Finally if $\sigma_2=\sigma_3$ and $\tau_2=\tau_3$ we obtain the generators $\omega_5$ and $\omega_6$:
\begin{equation}
\omega_5=\left(\begin{array}{cccc}
0&1&0&0\\
1&0&0&0\\
0&0&0&0\\
0&0&0&0
\end{array}\right),~~~
\omega_6=\left(\begin{array}{cccc}
0&0&0&0\\
0&0&0&0\\
0&0&0&-1\\
0&0&1&0
\end{array}\right)\, .
\end{equation}
We may identify the generators $\omega_1$, $\omega_3$ and $\omega_5$ as boost generators whereas $\omega_2$, $\omega_4$ and $\omega_6$ describe spatial rotations if $e_0$ is timelike. That this is indeed the case one can verify following the construction presented in section \ref{sec:24a}. In the three cases presented above the area metric has the local gauge group $o(1,1)\oplus o(2)$.\\
It is clear now that if $(\sigma_1,\tau_1)=(\sigma_2,\tau_2)=(\sigma_3,\tau_3)$, the area metric features full Lorentz symmetry and the local gauge algebra is $o(1,3)$.\\
Exactly along the same lines, one calculates the symmetry generators for the other metaclasses, depending on the possible degeneracies. This concludes the proof.\\

Finally, we remark that a direct calculation shows that an area metric that is induced by a metric with Lorentzian signature lies in metaclass I. This is not surprising since metaclass I area metrics are the only ones where the Lorentz group is one of the possible gauge groups. Similarly, one finds that area metrics induced by a Riemannian metric lie in class VII.

With the results of this section at hand, we now afford a complete algebraic overview over four-dimensional area metrics. The question whether this classification can be employed in deciding if an area metric manifold defines a spacetime is answered in the affirmative in the following section.

\section{Algebraic criteria for area metric spacetimes}
\label{chap:4}
As an application of the algebraic classification of four-dimensional area metrics, we discuss the various metaclasses with respect to their physical viability. The developments in the present section draw heavily on the causal structure of area metric spacetimes developed in section 2, and the algebraic classification obtained in section 3. In particular, we prove a theorem that excludes 16 of the 23 metaclasses of four-dimensional area metrics as viable spacetimes. An even stronger exclusion theorem is then obtained for spherically symmetric area metric spacetimes.

\subsection{Metaclasses containing no spacetimes}\label{sec:41}

With the explicit normal forms of an area metric at hand, we can discuss the families of area metrics contained in the various metaclasses with respect to their physical viability. We would like to know which area metrics provide possible spacetime backgrounds for dynamical systems such as Maxwell electrodynamics. The crucial ingredients to solve this question have been reviewed in section \ref{chap:2} and \ref{chap:3}. We now present a rather powerful theorem that excludes 16 of the 23 metaclasses as feasible backgrounds for physical theories since they do not provide strongly hyperbolic area metric spacetimes.

\begin{theorem}
\label{theo:41}
Let $(M,G)$ be a four-dimensional area metric manifold of metaclass VIII to XXIII. Then the Cauchy problem for Maxwell electrodynamics is not well-posed.
\end{theorem}

This theorem can be proven with the help of two lemmata. First note that $J^{-1}=\omega_GG^{-1}$ has the same Segr\'e-classification as $J$.
\begin{lemma}
\label{lem:41}
Let $(M,G)$ be a four-dimensional area metric manifold of metaclass VIII to XXIII. Then there exists a plane $\theta^1\wedge\theta^2$ of null covectors.
\end{lemma}
\textbf{Proof.} The endomorphism $J^{-1}$ has a real Jordan block of at least dimension two with eigenvalue $\lambda$. Then there exist $\Omega_1,\Omega_2\in \Lambda^2T_p^*M$ with $J^{-1}\Omega_1=\lambda\Omega_1$ and $J^{-1}\Omega_2=\Omega_1+\lambda\Omega_2$. Now $J^{-1}$ is symmetric with respect to the bilinear form $\omega_G^{-1}$. Expanding $\omega_G^{-1}(J^{-1}\Omega_1,\Omega_2)=\omega_G^{-1}(\Omega_1,J^{-1}\Omega_2)$ shows that $\omega_G^{-1}(\Omega_1,\Omega_2)=0$, and hence $\Omega_1$ is simple, i.e. there exist covectors $\theta_1$ and $\theta_2$ such that $\Omega_1=\theta_1\wedge\theta_2$. To show that this is a null plane consider $\xi\in\langle \theta^1,\theta^2\rangle$. Then we have $\xi\wedge J^{-1}(\theta^1\wedge\theta^2)=0$. Using the definition of $J^{-1}$ this may be rewritten as $G^{-1}(\xi,\cdot,\theta^1,\theta^2)=0$. This condition implies $\mbox{rk}~G^{-1}(\xi,\cdot,\xi,\cdot)<3$. By definition of the Fresnel polynomial (\ref{eqn:fresnelpoly}) it follows that all $\xi\in\langle \theta^1,\theta^2\rangle$ are null covectors: $\mathcal{G}(\xi,\xi,\xi,\xi)=0$, which concludes the proof.\\

We will use the result of this lemma in the proof of

\begin{lemma}
Let $(M,G)$ be a four-dimensional area metric manifold of metaclass VIII to XXIII. Then for every time space splitting there always exists a spatial covector $k$ such that the matrix $C(k)=C^{\alpha}k_{\alpha}$ defined in (\ref{eqn:Cauchy}) is not diagonalizable.
\end{lemma}
\textbf{Proof.} Choose an arbitrary time component $t$ with corresponding initial data surface of gradient $dt$. Without loss of generality assume that $\mathcal{G}(dt,dt,dt,dt)\not= 0$.\\
To show that there exists a covector $k$ such that the matrix $C(k)$ is not diagonalizable we compare the geometric and algebraic multiplicities of the zero eigenvalues of $C(k)$.\\
We choose the covector basis $(\theta^0=dt,\theta^1,\theta^2,\theta^3)$ where $\theta^1\wedge\theta^2$ is the discussed null plane. The eigenvalues $\lambda$ of the matrix $C(k)$ can be calculated according to
\begin{equation}
\mbox{det}(\lambda\mathds{1}-C^{\alpha}k_{\alpha})=\lambda^2\mathcal{G}(\lambda dt+k_{\alpha}dx^{\alpha},\dots,\lambda dt+k_{\alpha}dx^{\alpha})=0
\label{eqn:eigenv}
\end{equation}
for any spatial covector $k$. We now examine the particular covector $\theta^{\tilde{1}}=\theta^1+a\theta^2$ for some $a\not= 0$. Then equation (\ref{eqn:eigenv}) implies together with
\begin{equation}
\mathcal{G}(\lambda dt+\theta^{\tilde{1}},\dots,\lambda dt+\theta^{\tilde{1}})=\lambda^4\mathcal{G}^{0000}+4\lambda^3\mathcal{G}^{000\tilde{1}}+6\lambda^2\mathcal{G}^{00\tilde{1}\tilde{1}},
\end{equation}
that at least four of the six eigenvalues of the matrix $C(\theta^{\tilde{1}})$ are zero. Here we used that $\mathcal{G}^{\tilde 1\tilde 1\tilde 1\tilde 1}$ vanishes since $\theta^{\tilde 1}$ is a null covector and $\mathcal{G}^{0\tilde 1\tilde 1\tilde 1}=0$ since $G^{-1}(\xi,\cdot,\theta^1,\theta^2)=0$ for any $\xi\in \theta^1\wedge\theta^2$ as we have seen in the proof of lemma \ref{lem:41}. For the geometric multiplicity of the zero eigenvalues we have to find the number of eigenvectors $(u,v)^t$ corresponding to the zero eigenvalues of $C(\theta^{\tilde{1}})$. Finding the eigenvectors of $C(\theta^{\tilde{1}})$ corresponding to the zero eigenvalues is equivalent to solving the system $Ru=0$ and $Pu+Qv=0$ with matrices
\begin{equation}
R=\left[\begin{array}{ccc}
0&0&a\\
0&0&-1\\
-a&1&0
\end{array}\right],~~
P=\left[\begin{array}{ccc}
0&0&0\\
0&0&0\\
G^{2331}+aG^{2332}&G^{3131}+aG^{3132}&0
\end{array}\right],
\end{equation}
\begin{equation}
Q=\left[\begin{array}{ccc}
0&0&G^{0131}+2aG^{0(13)2}\\
0&0&2G^{0(23)1}+aG^{0232}\\
G^{0131}+2aG^{0(13)2}&2G^{0(23)1}+aG^{0232}&2G^{0331}+2aG^{0332}
\end{array}\right]\,.
\end{equation}
for vectors $u$ and $v$. We now observe that the choice of $\theta^{\tilde{1}}$ never gives rise to four eigenvectors, unless $u\in \langle u_0=(1,a,0)^t\rangle$, $Pu_0=0$ and $Q=0$. If the area metric is such that this cannot happen the proof is already complete. An additional step is only needed for area metrics with vanishing $G^{2331}$, $G^{2332}$, $G^{3131}$, $G^{0131}$, $G^{0232}$ and $G^{0231}=G^{0123}$. In this case we change the spatial covector to $\theta^{\tilde{1}}=\theta^1+ b\theta^3$. Along similar lines it can now be shown that the geometric multiplicity is always lower than the algebraic multiplicity. Hence, there always exists a spatial covector $k$ such that the matrix $C(k)$ is not diagonalizable. Since the coordinate choice of time $t$ was arbitrary this completes the proof.\\

From these two lemmata we immediately see that the proposition of theorem \ref{theo:41} holds. This theorem is quite a strong restriction on physically viable area metric backgrounds. That means we can restrict our further analysis of the normal forms to the physically viable metaclasses I to VII. Within these classes there may still be area metrics that do not admit a well-posed initial value problem for Maxwell electrodynamics. Since we have not been able to prove general theorems on metaclasses I to VII, their hyperbolicity properties have to be decided case by case. In the following section we present such a treatment for the case of spherically symmetric area metric manifolds.

\subsection{Highly symmetric area metric spacetimes}\label{sec:42}
Invariance of an area metric tensor field $G$ under its flow along some vector field $K$ identifies a symmetry of the area metric manifold and is conveniently formulated as a vanishing Lie derivative $\mathcal{L}_K G$, in complete analogy to pseudo-Riemannian geometry. Under the assumption of sufficiently high symmetry, we now further refine our study of the hyperbolicity properties of classes I to VII, on which the theorem proven in the previous chapter makes no statement.

In particular, we now examine spherically symmetric area metric spacetimes in some detail and comment on the simpler case of homogeneous and isotropic symmetry. We will see that the symmetries alone do not yet determine a unique metaclass. However, requiring that the area metric manifold is strongly hyperbolic will reveal that only metaclass I is physically viable.

To make these statements precise, note that the inverse $G^{-1}$ of some area metric tensor lies in the same metaclass as $G$, and we make the

\begin{definition}
An area metric manifold $(M,G)$ is called spherically symmetric if the area metric $G$ possesses three Killing vector fields spanning an $so(3)$ algebra such that the orbit of any point under the corresponding isometries is topologically a two-sphere.
\end{definition}
A slight modification of the calculation in \cite{Punzi:2007di} now reveals the canonical form of the inverse $G^{-1}$ of a spherically symmetric area metric $G$. In a suitable local covector frame $\{\theta^a\}$, the Petrov matrix $Petrov(G^{-1})$ takes the form
\begin{equation}
Petrov(G^{-1})=\left[\begin{array}{cccccc}
\xi&0&0&0&0&2\sigma+\tau\\
0&\epsilon_2&0&0&-\sigma+\tau&0\\
0&0&\epsilon_2&-\sigma+\tau&0&0\\
0&0&-\sigma+\tau&\epsilon_1&0&0\\
0&-\sigma+\tau&0&0&\epsilon_1&0\\
2\sigma+\tau&0&0&0&0&\epsilon_3^2
\end{array}\right]\, ,
\end{equation}
where $\xi$, $\sigma$ and $\tau$ are functions of $t$ and $r$, and the $\epsilon_1$, $\epsilon_2$, $\epsilon_3$ are signs of possible values $0$, $\pm 1$.
From the exclusion theorem 4.1 it is clear that not every combination of signs $\epsilon_1$, $\epsilon_2$, $\epsilon_3$ can possibly give rise to an area metric spacetime. The allowed combinations of signs are summarized in table \ref{tab:signs}.

\begin{table*}[h!]
	\centering
		\begin{tabular}{|c|c|c|c|c|c|}
			\hline
			$\epsilon_1$ & $\pm 1$&$\pm 1$ &$\pm 1$&$\pm 1$&0 \\
			\hline
			$\epsilon_2$ & $\mp 1$ & $\mp 1$ &$\pm 1$&$\pm 1$&0\\
			\hline
			$\epsilon_3^2$& $1$ & $1$ &$1$ &$1$ &$1$ \\
			\hline
			$\mbox{sign}(\xi)$ & $-$ & $+$ & $-$ & $+$ & $+$\\
			\hline
			metaclass & I & IV & VI & VII & VII\\
			\hline			
		\end{tabular}
		\caption{possible combination of signs for spherically symmetric area metrics}
		\label{tab:signs}
\end{table*}

We now calculate the Fresnel polynomial $\mathcal{G}(k,k,k,k)$ for some covector $k$ with components $k_i$, $i=t,r,\theta,\phi$. Up to a power of $\mbox{det}(Petrov(G))$ we obtain
\begin{equation}
\mathcal{G}(k,k,k,k)\sim\xi u^2+(\epsilon_1\epsilon_2+\xi\epsilon_3^2-9\sigma^2)uv+\epsilon_1\epsilon_2\epsilon_3^2v^2,
\end{equation}
where $u=\epsilon_2k_t^2+\epsilon_1k_r^2$ and $v=k_{\theta}^2+k_{\phi}^2$. Now observe that there can not be future timelike covectors $k$ if $\xi=0$, since then $\mathcal{G}(k,k,k,k)\sim v$. If $\xi\not= 0$ we may factorize the Fresnel polynomial into two real quadratic factors, $\mathcal{G}(k,k,k,k)\sim (u+\zeta^{+} v)(u+\zeta^- v)$ with
\begin{equation}
\zeta^{\pm}=\frac{1}{2\xi}\Big(\epsilon_1\epsilon_2+\xi -9\sigma^2\pm\sqrt{(\epsilon_1\epsilon_2+\xi -9\sigma^2)^2-4\epsilon_1\epsilon_2\xi}\Big).
\end{equation}
The Fresnel polynomial now has the form $\mathcal{G}(k,k,k,k)\sim((g^+)^{ab}k_ak_b)((g^-)^{ab}k_ak_b)$ for two inverse metrics $g^{\pm}=\mbox{diag}(\epsilon_2,\epsilon_1,\zeta^{\pm},\zeta^{\pm})$. For a weakly hyperbolic area metric spacetime both $g^+$ and $g^-$ need to have Lorentzian signature which requires that $\epsilon_1$ and $\epsilon_2$ have opposite sign. According to table \ref{tab:signs} this rules out the metaclasses VI and VII. Without loss of generality we assume $\epsilon_1=1$ and $\epsilon_2=-1$. One may now calculate the future timelike covector cone $C_p^*$ for the metaclasses I and IV. It turns out that the future timelike covector cone of metaclass IV is empty while $C_p^*$ for metaclass I is
\begin{equation}
C_p^*=\{k\in T_p^*M| -k_t^2+k_r^2+\zeta^-(k_{\theta}^2+k_{\phi}^2)<0\}.
\end{equation}
Thus we see that spherically symmetric area metrics spacetimes do only exist in metaclass I. The same result is obtained for homogeneous and isotropic manifolds in four dimensions \cite{Punzi:2006nx}. With these insights we conclude our demonstration of the various ways in which the algebraic classification of area metrics can be employed in order to decide on the hyperbolicity properties of area metric manifolds, and thus their ability to serve as a refined spacetime structure.

\section{Conclusions}

The key achievement of the present work is the identification of those four-dimension\-al area metric manifolds that qualify as viable spacetimes. The latter are distinguished by enabling causal evolution for classical matter fields in general, and at the very minimum for Maxwell theory. Indeed, remarkably much can be learnt from the application of standard constructions within the theory of partial differential equations to abelian gauge fields on an area metric manifold. The entire causal structure of an area metric manifold is revealed this way.

In this context, the central insight consists in the observation that the naturally emerging future timelike cones are open and convex, and that their topological closure defines causal vectors. Based on these notions, we were able to provide analytical definitions for weakly and strongly hyperbolic area metric spacetimes, such that the known theorems concerning the well-definition of initial value problems directly extend from the familiar special case of metric manifolds. Indeed, we were able to rigorously develop all concepts needed to address global causality conditions, leading for instance to the area metric version of the equivalence of the Alexandrov topology with that of the underlying manifold whenever the area metric spacetime is strongly causal.

The second major technical part of this article, namely the complete algebraic classification of four-dimensional area metric manifolds, was prompted by the desire to obtain simple \textit{algebraic} criteria for the above \textit{analytic} characterization of strongly hyperbolic area metric spacetimes. Since four-dimen\-sional area metric manifolds contain more algebraic degrees of freedom at each point than could possibly be trivialized by a change of the local frame, the algebraic classification results in continuous families of normal forms. Grouping these into 23 metaclasses allows to prove a remarkable theorem, linking our analytical conditions for a strongly hyperbolic area metric spacetime to the obtained algebraic classification: 16 of the 23 metaclasses of area metrics cannot provide strongly hyperbolic spacetime geometries.

We wish to emphasize that currently we have comparatively little to say about the hyperbolicity properties of the remaining seven algebraic metaclasses, unless further assumptions, such as the existence of sufficiently many Killing symmetries, are made. That this does not necessarily present a problem in practice, we demonstrated by scrutinizing spherically symmetric area metric spacetimes as a concrete example of phenomenological interest. Here we were indeed able to give a full algebraic classification of all strongly hyperbolic spherically symmetric area metric spacetimes. The same holds for homogeneous and isotropic area metric spacetimes. In both cases, strong hyperbolicity is equivalent to the respective area metrics being of algebraic metaclass I.

It is interesting to briefly muse on what we have learnt beyond the immediate technical details when studying the physical viability of an area metric spacetime structure.

First and foremost, the questions discussed here for the particular case of area metric manifolds must be posed for any candidate geometry aspiring to replace the Lorentzian spacetime structure underlying general relativity and our current fundamental theories of matter. That indeed area metric geometry passes key criteria one must expect a spacetime geometry to satisfy provides further evidence toward the viability of the area metric hypothesis at a fundamental level.

Second, the treatment given here immediately includes the corresponding findings in the metric case, in which all constructions recover what is often merely postulated, but rarely emphasized to be intimately linked to other assumptions made in the theory. A case in point is the physically somewhat incomplete (though mathematically elegant) discussion of the causal structure of spacetime purely in terms of the geometry, but without pointing out the relation to (and indeed logical origin in) the hyperbolicity properties of distinguished matter fields. Thus, what might appear to be a more intricate treatment of these questions in area metric geometry actually only highlights the conceptual steps to be followed also in the familiar metric case.

Third, area metric spacetimes provide a now well-understood example for a geometry where local Lorentz invariance may be gradually broken (see section \ref{sec:34}) while, and this is a mathematically and physically important point, their causal structure is still given by convex cones. It is this latter property, which ultimately renders for example the decay of massless particles into massive ones kinematically impossible.

Naturally, the developments of the present article are of greatest value particularly for the further pursuit of the programme to study the physical implications of an area metric spacetime structure. The future cones, for instance, are of central relevance in defining local observers, and thus for the extraction of physical predictions from the theory. The normal forms, and particularly those that could be identified as providing strongly hyperbolic backgrounds, are of obvious value for actual calculations, and useful for a number of constructions that are not possible for non-hyperbolic area metric spacetimes. Work that has been enabled by these and other results obtained in this article is currently under way.

\paragraph*{Acknowledgments}
The authors wish to thank Gary Gibbons for most valuable comments and providing extensive notes on the geometry of convex cones, as well as Hans-Peter Tuschik for coming up with the crucial idea underlying the proof of the determinant identity in section \ref{sub:21}. The early stages of this work owe much to the contributions of Raffaele Punzi, before his untimely death. MNRW gratefully acknowlegdes support through the Emmy Noether Fellowship grant WO 1447/1-1. CW thanks the Studienstiftung des deutschen Volkes for their financial and ideal support through their academic programme, where this work with FPS was initiated.




\bibliography{bibliography}

\begin{thebibliography}{10}
\expandafter\ifx\csname url\endcsname\relax
  \def\url#1{\texttt{#1}}\fi
\expandafter\ifx\csname urlprefix\endcsname\relax\def\urlprefix{URL }\fi
\expandafter\ifx\csname href\endcsname\relax
  \def\href#1#2{#2} \def\path#1{#1}\fi

\bibitem{Beem}
J.~K. Beem, P.~E. Ehrlich, K.~L. Easley, Global Lorentzian Geometry, 2nd
  Edition, Marcel Dekker, 1996.

\bibitem{Wigner}
E.~P. Wigner, {On unitary representations of the inhomogenous Lorentz group},
  Ann. Math. 40 (1939) 149--204.

\bibitem{EPP}
P.~Ramond, Field theory: A modern primer, 2nd Edition, Frontiers in Physics,
  Perseus, 1997.

\bibitem{Einstein}
A.~Einstein, {Zur Elektrodynamik bewegter K\"orper}, Ann. Phys. (Leipzig) 4
  (1905) 891--921.

\bibitem{Kuchar}
S.~Hojman, K.~Kuchar, C.~Teitelboim, {Geometrodynamics Regained}, Ann. Phys.
  (N.Y.) 96 (1976) 88.

\bibitem{MTW}
C.~W. Misner, K.~S. Thorne, J.~A. Wheeler, Gravitation, 2nd Edition, W. H.
  Freeman, 1973.

\bibitem{Spergel}
D.~N. Spergel, et~al., {Wilkinson Microwave Anisotropy Probe (WMAP) three year
  results: Implications for cosmology}, Astrophys. J. Suppl. 170 (2007) 377.
\newblock \href {http://arxiv.org/abs/astro-ph/0603449}
  {\path{arXiv:astro-ph/0603449}}, \href {http://dx.doi.org/10.1086/513700}
  {\path{doi:10.1086/513700}}.

\bibitem{Knop}
R.~A. Knop, et~al., {New Constraints on $\Omega_M$, $\Omega_\Lambda$, and w
  from an Independent Set of Eleven High-Redshift Supernovae Observed with
  HST}, Astrophys. J. 598 (2003) 102.
\newblock \href {http://arxiv.org/abs/astro-ph/0309368}
  {\path{arXiv:astro-ph/0309368}}, \href {http://dx.doi.org/10.1086/378560}
  {\path{doi:10.1086/378560}}.

\bibitem{DEMrev}
V.~Sahni, {Dark matter and dark energy}, Lect. Notes Phys. 653 (2004) 141--180.
\newblock \href {http://arxiv.org/abs/astro-ph/0403324}
  {\path{arXiv:astro-ph/0403324}}.

\bibitem{Laemmerzahl}
C.~Laemmerzahl, O.~Preuss, H.~Dittus, {Is the physics within the Solar system
  really understood?}\href {http://arxiv.org/abs/gr-qc/0604052}
  {\path{arXiv:gr-qc/0604052}}.

\bibitem{newEPP}
A.~Boyarsky, A.~Neronov, O.~Ruchayskiy, M.~Shaposhnikov, {Constraints on
  sterile neutrino as a dark matter candidate from the diffuse X-ray
  background}, Mon. Not. Roy. Astron. Soc. 370 (2006) 213--218.
\newblock \href {http://arxiv.org/abs/astro-ph/0512509}
  {\path{arXiv:astro-ph/0512509}}, \href
  {http://dx.doi.org/10.1111/j.1365-2966.2006.10458.x}
  {\path{doi:10.1111/j.1365-2966.2006.10458.x}}.

\bibitem{MOND}
J.~D. Bekenstein, {Relativistic gravitation theory for the MOND paradigm},
  Phys. Rev. D70 (2004) 083509.
\newblock \href {http://arxiv.org/abs/astro-ph/0403694}
  {\path{arXiv:astro-ph/0403694}}, \href
  {http://dx.doi.org/10.1103/PhysRevD.70.083509}
  {\path{doi:10.1103/PhysRevD.70.083509}}.

\bibitem{Punzi:2006hy}
R.~Punzi, F.~P. Schuller, M.~N.~R. Wohlfarth, {Geometry for the accelerating
  universe}, Phys. Rev. D76 (2007) 101501.
\newblock \href {http://arxiv.org/abs/hep-th/0612133}
  {\path{arXiv:hep-th/0612133}}, \href
  {http://dx.doi.org/10.1103/PhysRevD.76.101501}
  {\path{doi:10.1103/PhysRevD.76.101501}}.

\bibitem{Schuller:2005yt}
F.~P. Schuller, M.~N.~R. Wohlfarth, {Geometry of manifolds with area metric},
  Nucl. Phys. B747 (2006) 398--422.
\newblock \href {http://arxiv.org/abs/hep-th/0508170}
  {\path{arXiv:hep-th/0508170}}, \href
  {http://dx.doi.org/10.1016/j.nuclphysb.2006.04.019}
  {\path{doi:10.1016/j.nuclphysb.2006.04.019}}.

\bibitem{Drummond:1979pp}
I.~T. Drummond, S.~J. Hathrell, {QED Vacuum Polarization in a Background
  Gravitational Field and Its Effect on the Velocity of Photons}, Phys. Rev.
  D22 (1980) 343.
\newblock \href {http://dx.doi.org/10.1103/PhysRevD.22.343}
  {\path{doi:10.1103/PhysRevD.22.343}}.

\bibitem{Schuller:2005ru}
F.~P. Schuller, M.~N.~R. Wohlfarth, {Canonical differential structure of string
  backgrounds}, JHEP 02 (2006) 059.
\newblock \href {http://arxiv.org/abs/hep-th/0511157}
  {\path{arXiv:hep-th/0511157}}.

\bibitem{Punzi:2009ks}
R.~Punzi, F.~P. Schuller, M.~N.~R. Wohlfarth, {Light clocks in strong
  gravitational fields}\href {http://arxiv.org/abs/0902.1811}
  {\path{arXiv:0902.1811}}.

\bibitem{Punzi:2006nx}
R.~Punzi, F.~P. Schuller, M.~N.~R. Wohlfarth, {Area metric gravity and
  accelerating cosmology}, JHEP 02 (2007) 030.
\newblock \href {http://arxiv.org/abs/hep-th/0612141}
  {\path{arXiv:hep-th/0612141}}.

\bibitem{BDgeometry}
R.~Punzi, F.~P. Schuller, M.~N.~R. Wohlfarth, {Brans-Dicke geometry}, Phys.
  Lett. B670 (2008) 161--164.
\newblock \href {http://arxiv.org/abs/0804.4067} {\path{arXiv:0804.4067}},
  \href {http://dx.doi.org/10.1016/j.physletb.2008.10.046}
  {\path{doi:10.1016/j.physletb.2008.10.046}}.

\bibitem{Tuschik}
H.-P. Tuschik, private communication.

\bibitem{Symplectic}
A.~Cannas Da~Silva, Lectures on symplectic geometry, Lecture Notes in
  Mathematics, Springer, 2001.

\bibitem{Cartan}
E.~Cartan, Les espace métriques fondés sur la notion d'aire, Hermann, Paris,
  1933.

\bibitem{Polch:1}
J.~Polchinski, String theory, Volume I: An Introduction to the bosonic string,
  Camebridge University Press, 1998.

\bibitem{Polyakov}
A.~M. Polyakov, {Quantum geometry of bosonic strings}, Phys. Lett. B103 (1981)
  207--210.
\newblock \href {http://dx.doi.org/10.1016/0370-2693(81)90743-7}
  {\path{doi:10.1016/0370-2693(81)90743-7}}.

\bibitem{Einstein:1}
A.~Einstein, The Meaning of Relativity, 5th Edition, Princeton University
  Press, 2004.

\bibitem{Papapetrou:1948jw}
A.~Papapetrou, {Einstein's theory of gravitation and flat space}, Proc. Roy.
  Irish Acad. (Sect. A) 52A (1948) 11--23.

\bibitem{Moffat:1995fc}
J.~W. Moffat, {Nonsymmetric gravitational theory}, J. Math. Phys. 36 (1995)
  3722--3732.
\newblock \href {http://dx.doi.org/10.1063/1.530993}
  {\path{doi:10.1063/1.530993}}.

\bibitem{Punzi:2009yq}
R.~Punzi, F.~P. Schuller, M.~N.~R. Wohlfarth, {Massive motion in area metric
  spacetimes}, Phys. Rev. D 79 (2009) 124025.
\newblock \href {http://arxiv.org/abs/0901.3264} {\path{arXiv:0901.3264}}.

\bibitem{Punzi:2007di}
R.~Punzi, M.~N.~R. Wohlfarth, F.~P. Schuller, {Propagation of light in area
  metric backgrounds}, Class. Quant. Grav. 26 (2009) 035024.
\newblock \href {http://arxiv.org/abs/0711.3771} {\path{arXiv:0711.3771}},
  \href {http://dx.doi.org/10.1088/0264-9381/26/3/035024}
  {\path{doi:10.1088/0264-9381/26/3/035024}}.

\bibitem{Laemmerzahl2}
C.~Laemmerzahl, The geometry of matter fields, in: V.~de~Sabbata, J.~Audretsch
  (Eds.), Quantum mechanics in curved spacetime.

\bibitem{Wald}
R.~M. Wald, General Relativity, 1st Edition, The University of Chicago Press,
  Chicago and London, 1984.

\bibitem{PDE:1}
S.~Benzoni-Gavage, D.~Serre, Multi-dimensional Hyperbolic Partial Differential
  Equations: First-order Systems and Applications, Oxford University Press,
  2007.

\bibitem{Rubilar}
G.~F. Rubilar, {Linear pre-metric electrodynamics and deduction of the light
  cone}, Annalen Phys. 11 (2002) 717--782.
\newblock \href {http://arxiv.org/abs/0706.2193} {\path{arXiv:0706.2193}}.

\bibitem{Hehlbook}
F.~W. Hehl, Y.~N. Obukhov, Foundations of classical electrodynamics,
  Birkh\"auser, Basel, 2003.

\bibitem{Garding1}
L.~G\r{a}rding, {Linear hyperbolic equations with constant coefficients}, Acta
  Math. 85 (1951) 1--62.

\bibitem{Renegar}
J.~Renegar, {Hyperbolic Programs, and Their Derivative Relaxations},
  Foundations of Computational Mathematics 6 (2006) 59.

\bibitem{Garding2}
L.~G\r{a}rding, {An inequality for hyperbolic polynomials}, J. Math. 8 (1959)
  957--965.

\bibitem{Renegar2}
J.~Renegar, {A mathematical view of interior-point methods in convex
  optimization}, SIAM.

\bibitem{Lang}
S.~Lang, Linear Algebra, Springer, 2004.

\bibitem{Reisenberger:1995xh}
M.~P. Reisenberger, {New constraints for canonical general relativity}, Nucl.
  Phys. B457 (1995) 643--687.
\newblock \href {http://arxiv.org/abs/gr-qc/9505044}
  {\path{arXiv:gr-qc/9505044}}, \href
  {http://dx.doi.org/10.1016/0550-3213(95)00448-3}
  {\path{doi:10.1016/0550-3213(95)00448-3}}.

\bibitem{DePietri:1998mb}
R.~De~Pietri, L.~Freidel, {so(4) Plebanski Action and Relativistic Spin Foam
  Model}, Class. Quant. Grav. 16 (1999) 2187--2196.
\newblock \href {http://arxiv.org/abs/gr-qc/9804071}
  {\path{arXiv:gr-qc/9804071}}, \href
  {http://dx.doi.org/10.1088/0264-9381/16/7/303}
  {\path{doi:10.1088/0264-9381/16/7/303}}.

\bibitem{Algebra:1}
I.~Gohberg, P.~Lancaster, L.~Rodman, Indefinite Linear Algebra and
  Applications, Birkhäuser, 2005.

\bibitem{Segre}
G.~S. Hall, Symmetries and curvature structure in general relativity, World
  Scientific, 2004.

\end{thebibliography}
\bibliographystyle{elsarticle-num}

\end{document}